\documentclass[aps,twocolumn,prx,amsfonts,showpacs,superscriptaddress,longbibliography]{revtex4-2}

\usepackage{graphicx}
\usepackage{bm}
\usepackage{hyperref}
\usepackage{amsmath}
\usepackage{amssymb}
\usepackage[table]{xcolor}
\usepackage{array}
\usepackage{multirow}
\usepackage[caption=false]{subfig}
\usepackage{physics}
\usepackage{soul}

\usepackage{amsthm}
\usepackage{bbm}
\usepackage{multirow}
\usepackage[colorinlistoftodos]{todonotes}
\usepackage[scr=boondoxo,frak=boondox,scrscaled=1.05]{mathalfa}
\usepackage{amsmath}

\def\ket#1{|#1\rangle}
\def\braket#1#2{\langle#1|#2\rangle}

\def\r{{\boldsymbol{r}}}

\def\k{{\boldsymbol{k}}}
\def\p{{\boldsymbol{p}}}

\def\q{{\boldsymbol{q}}}

\def\G{{\boldsymbol{G}}}

\def\a{{\boldsymbol{a}}}
\def\b{{\boldsymbol{b}}}

\newcommand{\bq}{\bm {q}}

\newcommand{\bk}{\bm {k}}
\newcommand{\bp}{\bm {p}}

\newcommand{\br}{\bm {r}}
 
\usepackage{ulem}
\begin{document}

\title{From Fractionalization to Chiral Topological Superconductivity  in a Flat Chern Band}

\author{Daniele Guerci}
\thanks{These authors contributed equally to this work.}
\affiliation{Department of Physics, Massachusetts Institute of Technology, Cambridge, MA-02139, USA}

\author{Ahmed Abouelkomsan}
\thanks{These authors contributed equally to this work.}
\affiliation{Department of Physics, Massachusetts Institute of Technology, Cambridge, MA-02139, USA}

\author{Liang Fu}
\affiliation{Department of Physics, Massachusetts Institute of Technology, Cambridge, MA-02139, USA}

\begin{abstract}

We show that interacting electrons in a flat Chern band can form, in addition to fractional Chern insulators,  a chiral $f$-wave topological superconductor that hosts neutral Majorana fermion edge modes. Superconductivity emerges from an interaction-induced metallic state that exhibits  anomalous Hall effect, as observed in rhombohedral graphene and near the $\nu=\frac{2}{3}$ fractional Chern insulator in twisted transition metal dichalcogenides.

\end{abstract}

\maketitle

\date{\today}

{\it Introduction---} The recent observation of fractional quantum anomalous Hall (FQAH) effect in moir\'e materials \cite{Cai2023,zeng2023thermodynamic, xuParkObservationFractionallyQuantized2023, PhysRevX.13.031037,Lu2024Feb} has reinforced interest in flat topological bands as a promising venue for realizing novel quantum matter. 
As a common feature, these materials host conjugate Chern bands with opposite Chern numbers in $K$ and $K'$ valleys, which form time-reversed pairs \cite{FW_PRL_2019,Devakul_2021}. 
While narrow bandwidth generally enhances interaction effects, the nontrivial Bloch wave functions in topological bands drive interesting physics beyond the canonical Hubbard model.   

First, Coulomb interaction tends to drive spontaneous ferromagnetism and lift the degeneracy between conjugate Chern bands. Importantly, ferromagnetism occurs not only at integer band fillings \cite{Devakul_2021,lian2021twisted,bultinck2020ground,bultinck2020mechanism,repellin_ferromagnetism_2020,wu2020collective}, but extends over a {\it wide} and {\it continuous} range of partial band fillings \cite{crepel2023anomalous,reddyFQAH,Anderson2023}. This enables the FQAH effect to occur at fractional fillings when spin- and valley-polarized electrons populate a {\it single} Chern band and form a fractional Chern insulator (FCI) as theoretically predicted for twisted transition metal dicholcogenides \cite{crepel2023anomalous,Li2021PRR}. 

Notably, the FCI states observed so far exhibit the same quantized Hall plateaus as the celebrated Jain sequence of fractional quantum Hall states. This fact suggests some level of similarity between the underlying Chern band and the lowest Landau level. However, unlike Landau levels, a generic Chern band has nonuniform  Berry curvature and quantum metric in the Brillouin zone, which can be experimentally tuned for example by the twist angle~\cite{Devakul_2021,Tarnopolsky_2019,Nicolas2023_prr,AllanNicolas_2024}. It is therefore interesting to explore the possibility of novel phases in partially filled Chern bands that do not have a Landau level analog, when the  band geometry is tuned away from the Landau level limit~\cite{Abouelkomsan_2023,reddy2023toward}.

In this work, we theoretically discover chiral superconductivity from  purely repulsive electron interaction in flat Chern bands, even when electron's kinetic energy is {\it completely} quenched in a {\it dispersionless} band. Our superconductor has $f$-wave pairing symmetry and is a time-reversal-breaking topological state hosting $N=3$ branches of chiral Majorana fermion  edge modes. Remarkably, we find that superconductivity and fractionalization can coexist in the same system, either at different filling factors, or at the same filling factor with the transition between the two tuned by the geometry of the underlying Chern band. 

Our findings are obtained from a combination of analytical and numerically exact many-body calculation. 
Our Chern band model is motivated by and relevant to twisted transition metal dichalcogenides. Intriguing signatures of chiral superconductivity have been observed in rhombohedral graphene in the absence of FQAH effect~\cite{han2025signatureschiralsuperconductivityrhombohedral}. Also, a very recent work reported superconductivity in twisted MoTe$_2$ that coexists with the FQAH state at band filling $\nu=\frac{2}{3}$ ~\cite{xu2025signaturesunconventionalsuperconductivitynear}. 

\begin{figure}
    \centering
    \includegraphics[width=\linewidth]{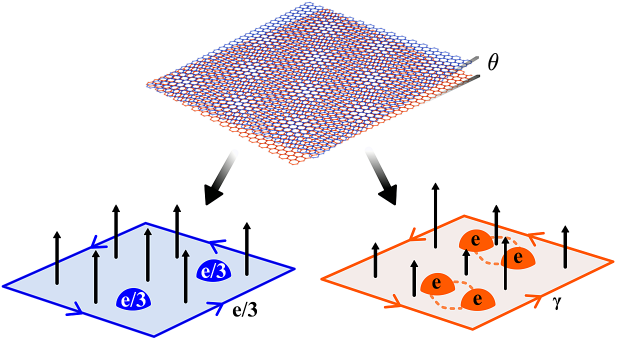}
    \caption{A schematic: electrons in twisted semiconductors in the absence of magnetic fields can be mapped to particles in a  periodically varying magnetic field. Fractional Chern insulators (left) are favored when the emergent magnetic field is nearly uniform. On the other hand, the magnetic field inhomogeneity drives chiral topological superconductivity with Majorana fermion edge modes (right).   }
    \label{fig:enter-label}
\end{figure}

Our Letter offers a proof of principle for chiral superconductivity in flat Chern band, and more importantly, reveals the key role of band geometry in tuning the competition between superconductivity and fractionalization. As the band geometry can be experimentally tuned by the twist angle and displacement field, our findings serve as a useful guide to ongoing and future experimental studies.   

{\it The model---} 
To start, we consider electrons with {\it repulsive} interaction in a perfectly flat Chern band, which is the most adverse setting for superconductivity by conventional wisdom. Assuming that the interaction scale is smaller than the gap to remote bands, the band-projected Hamiltonian consists of two-body interaction terms only, taking the general form   
\begin{equation}\label{hamiltonian}
     % H=\frac{1}{2A}\sum_{\bk_1\cdots \bk_4}H_{\bk_1,\bk_2;\bk_3,\bk_4}\, c^\dagger_{\bk_1 }c^\dagger_{\bk_2}c_{\bk_3}c_{\bk_4},
    H=\frac{1}{2A}\sum_{\q}V({\q}):\bar\rho({\q})\bar\rho({-\q}):,
\end{equation}
where $\bar\rho({\q})=\sum_{\k\in \rm BZ}\braket{u_\k}{u_{\k+\q}}c^\dagger_{\k}c_{\k+\q}$ is the projected density operator with $c^\dagger_\k$ creating a particle at crystal momentum $\bk$; note that $c^\dagger_\k = \eta_{\G} e^{-i\G\times\k\ell^2_B/2} c^\dagger_{\k+\G}$ for any reciprocal vector $\G$ and $\eta_{\G}$ parity of $\G$.  
Here, $u_{\k}$ is the cell periodic part of the Bloch wave function; $``:\,:"$ denotes normal ordering with respect to the empty band; and $V(\q)$ is the Fourier transform of the interaction potential $V(\r_1-\r_2)$, with $\q\in(-\infty, \infty)$.

For a given interaction potential, the Hamiltonian is entirely determined by the form factor ${\Lambda}_{\k,\k+\q} \equiv \braket{u_\k}{u_{\k+\q}}$, which depends only on the band wave function in $\k$ space. Notably, the form factor encodes the band geometry: at small $\q$, $|\Lambda_{\k,\k+\q}|$ is directly related to the quantum metric $g(\k)$, while the Berry curvature $\Omega(\bk)$ is encoded in the phase of the gauge-invariant product of form factors~\cite{Vanderbilt_2018,Fukui_2005}.

Both the quantum metric and Berry curvature are constant in the case of Landau levels, but not in generic Chern bands. To study new physics due to nonuniform band geometry, we consider a minimal generalization of the lowest Landau level wave function which describes a modulated Landau level produced by a spatially varying, periodic magnetic field~\cite{Dubrovin_1980,Aharonov_1979,dong2022diracelectronperiodicmagnetic}: 
\begin{equation}\label{app:C_1_idealband}
    \psi_{\bk}(\br)=\mathcal N_{\bk} e^{-K(\br)} \Phi_{\bk}(\br),
\end{equation}
where $\Phi_{\bk}(\br)$ denotes the usual $n=0$ Landau level wave function in magnetic Bloch basis, and $\mathcal N_{\bk}$ is the normalization factor. Importantly, $K(\br)$ is a periodic function of $\r$, which modulates the charge density of the Chern band in real space and tunes its band geometry in $\bk$ space~\cite{wang2021,Ledwith2020,Ledwith2023,guerci2024layerskyrmionsidealchern}. The Chern band reduces to the lowest Landau level when $K(\br)$ is constant, in which case the Berry curvature is $\k$-independent due to continuous magnetic translation symmetry, and departs significantly from it when $K(\br)$ is highly nonuniform.

\begin{figure}
    \centering
    \includegraphics[width=0.9\linewidth]{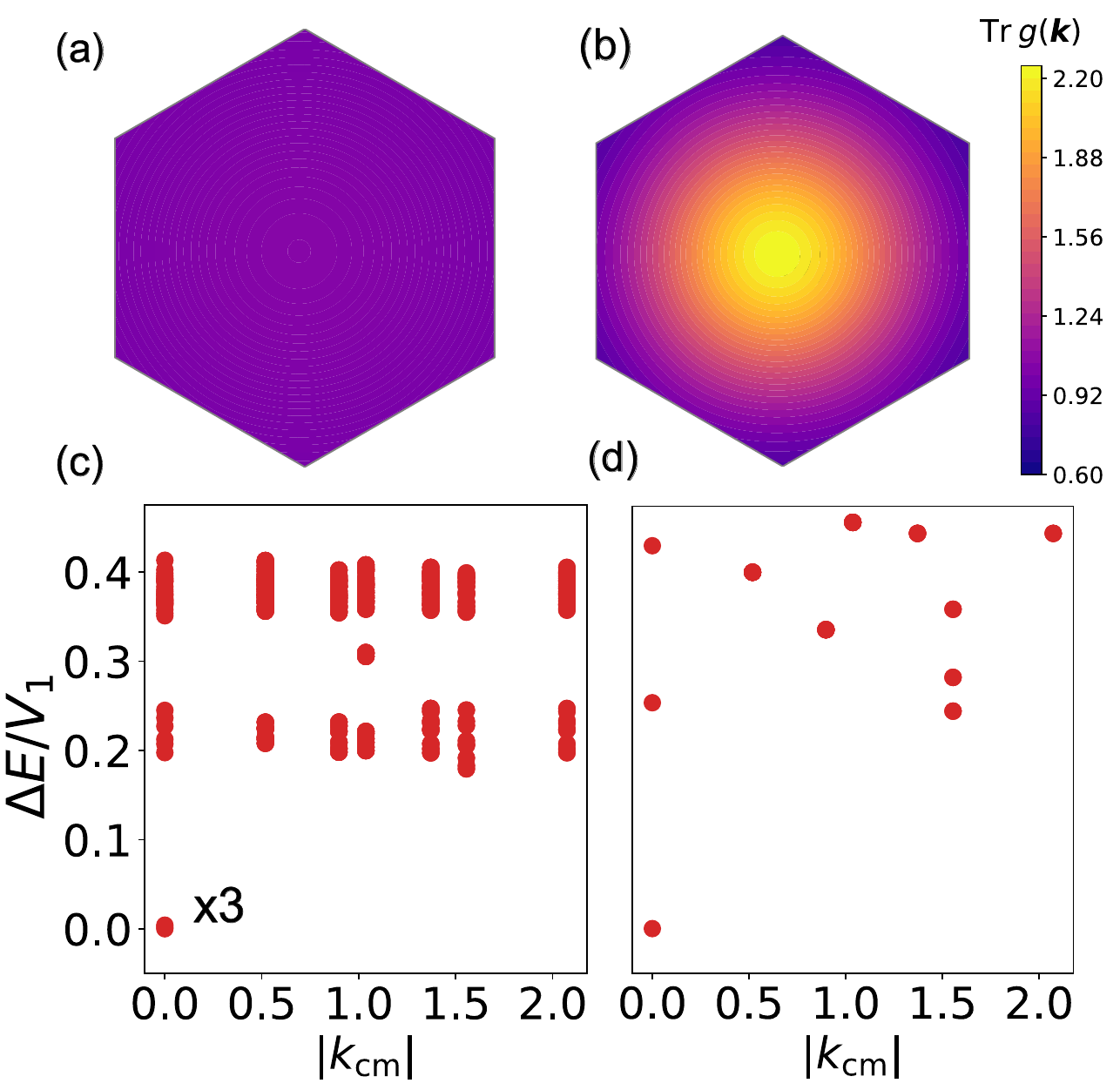}
    \caption{(a-b) The quantum metric trace ${\rm Tr} g(\k)$ in units such that the average is unity for (a) $\mathcal K = 0.05$ and (b) $\mathcal K = 0.8$ (Eq. \eqref{eq:Kofr}). (c-d) The many-body spectrum (in units of $V_1$) obtained from exact diagonalization as a function of the absolute value of the total momentum $|\k_{\rm cm}|$ for (c) $\mathcal K = 0.05$  and (d) $\mathcal K = 0.8$. Calculations were done on a 27 site cluster. }
    \label{fig:instabilityFAQH}
\end{figure}

Interestingly, this Chern band model naturally describes the lowest band of twisted transition metal dichalcogenides ($t$TMDs) to a good approximation. Here, an effective magnetic field emerges from the layer pseudospin texture which is spatially varying with the moir\'e periodicity~\cite{Paul_2023,Shi_2024,wu2025}. 
Motivated by its connection with $t$TMDs, in the following we model $K(\r)$ as a sum of the lowest harmonics reciprocal lattice vectors,  
\begin{equation}
\label{eq:Kofr}
K(\r) = -\dfrac{\sqrt{3}}{4\pi} \mathcal K  \sum_{i = 1,2,3}\cos(\b_i \cdot \r)     
\end{equation} 
with $\b_1=4\pi/(\sqrt{3}a)(1/2,\sqrt{3}/2)$, $\b_2=-4\pi/(\sqrt{3}a)(1,0)$ and $\b_3=-\b_1-\b_2$. $a$ is the underlying lattice constant.

\begin{figure*}
    \centering
    \includegraphics[width=0.8 \linewidth]{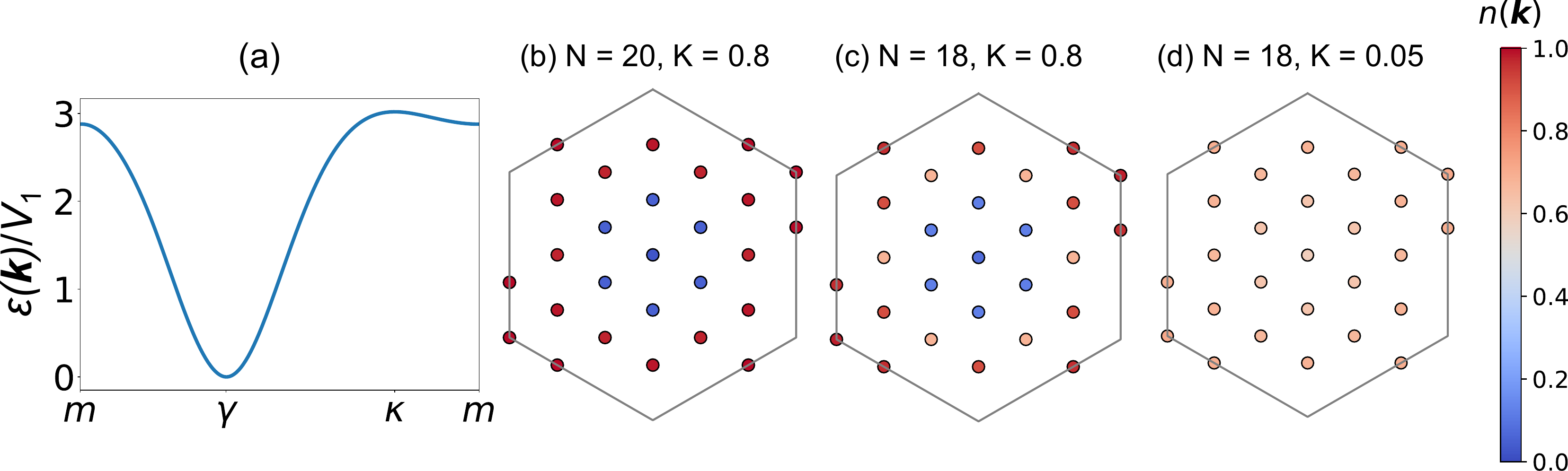}
    \caption{(a) Interaction-induced dispersion $\epsilon(\k)$ (Eq. \eqref{Hamiltonian_PH}) (in units of $V_1$) for $\mathcal K = 0.8$ plotted along a cut in the Brillouin zone. (b-d) Momentum occupation $n(\k) = \langle c^{\dagger}_{\k} c_{\k} \rangle$ evaluated in the many-body ground state.}
    \label{fig:dispersion}
\end{figure*} 

We first study the Hamiltonian \eqref{hamiltonian} with exact diagonalization (ED), $V(\r) = (3 V_1 a^4/ 4 \pi) \nabla^2 \delta(\r)$, i.e., $V(\q) = -v_1 \q^2$ with $v_1=3 V_1 a^4/ 4 \pi$.
For this short-range repulsion, previous works show that a generalized Laughlin wave function is the exact zero-energy ground state at $\nu =\frac{1}{3}$, regardless of the form of $K(\br)$ \cite{Ledwith2020,wang2021,dong2022diracelectronperiodicmagnetic, Ledwith2023}. 
On the other hand, as we show below, the situation is completely different at $\nu =\frac{2}{3}$ where the  ground state depends crucially on the geometry of the underlying Chern band, which is controlled by the parameter $\mathcal K$ in our model.

Figs.~\ref{fig:instabilityFAQH}(a) and~\ref{fig:instabilityFAQH}(b) show the trace of quantum metric for our Chern band at two different values of $\mathcal K$: it is nearly uniform for small $\mathcal K$, but highly nonuniform and peaked at the $\gamma$ point for large $\mathcal K$.  
The corresponding many-body energy spectra at $\nu = \frac{2}{3}$ are shown in Figs.~\ref{fig:instabilityFAQH}(c) and~\ref{fig:instabilityFAQH}(d), respectively.  
For nearly uniform band geometry, Fig.\ref{fig:instabilityFAQH}(c) shows that the many-body ground state remains a Laughlin-like FCI, as evidenced by the characteristic threefold topological degeneracy on the torus. However, as the quantum metric fluctuation increases~\cite{onishi2025geometricboundstructurefactor}, we find a new phase characterized by a single ground state at the $\gamma$ point in the BZ, see Fig.~\ref{fig:instabilityFAQH} (d).

As shown above,  the inhomogeneous band geometry at large $\mathcal K$ leads to a new phase at $\nu=\frac{2}{3}$,  but not at $\nu=\frac{1}{3}$. 
The contrasting behavior between $\frac{1}{3}$ and $\frac{2}{3}$ fillings is allowed because particle-hole symmetry is generally absent at $\mathcal K\neq 0$. As we show below, understanding the particle-hole asymmetry holds the key to unlocking the phase diagram of interacting electrons in flat Chern bands.

To proceed, we perform the particle-hole (PH) transformation $c_{\k} \equiv d^\dagger_{-\k}$ and rewrite the projected Hamiltonian $H$ in terms of hole operators: 
\begin{equation}
\label{Hamiltonian_PH}
    H =  \sum_\k \epsilon (\k)d^\dagger_{\k} d_{\k}+\frac{1}{2A}\sum_{\q}V({\q}):\bar{n}({\q})\bar n({-\q}):,
\end{equation}
up to a constant, where $\bar{n}(\q)=\sum_{\k}\Lambda^*_{-\k,-\k-\q} d^\dagger_{\k} d_{\k+\q}$ is the projected density operator of holes. 
Our system of electrons at filling factor $\nu$ can be equally viewed as that of holes at filling factor $1-\nu$ described by the Hamiltonian~\eqref{Hamiltonian_PH}. 
Importantly,  this hole Hamiltonian now contains an interaction-induced one-body term $\epsilon(\k)$, which is constant in the special case of $n=0$ Landau level but {\it dispersive} for {\it generic} Chern band systems, as detailed in the Supplemental Material (SM)~\cite{supplementary}, which includes Refs.~\cite{Giuliani_Vignale_2005,Ghazaryan_2021}.
In other words, holes acquire an interaction-induced energy dispersion  \cite{lauchli_hierarchy_2013,abouelkomsan_particle-hole_2020, Abouelkomsan_2023, liu_broken_2024, reddy2023toward}.

In our Chern band model with $\mathcal K \neq  0$, the interaction-induced hole dispersion $\epsilon(\k)$ is shown in Fig. \ref{fig:dispersion}(a), which has a single minima at the $\gamma$ point and a bandwidth on the order of interaction energy (which is the only energy scale in the problem). It is instructive to note that in the small-momentum transfer limit,  $\epsilon(\bk) \approx \frac{1}{A} \sum_{\q} V(\q)[1 - q_a q_b g_{ab}(\k)]$,  %with $a,b = x,y$. In this limit, 
i.e., the interaction-induced hole dispersion is entirely governed by the fluctuation of  quantum metric over the Brillouin zone. Indeed, comparing Fig. \ref{fig:instabilityFAQH}(b) and Fig. \ref{fig:dispersion}(a), we find that qualitative features of $\epsilon(\k)$ are accurately captured by the quantum metric. 

The exact transformation from electron to hole description offers a controlled method to study flat Chern band systems at band filling $\nu = 1 - \delta$ with $\delta \ll 1$. Such a nearly filled Chern band is naturally described by a small concentration $\delta$ of holes, which are \textit{weakly} interacting because the average interparticle distance is large compared to the interaction range. %Despite the strong-coupling nature of our system, 
Therefore, the ground state of generic flat Chern bands at band fillings close to $\nu=1$ is an emergent Fermi liquid of holes, in which the kinetic energy of holes dominates over their residual interaction. %This conclusion holds for generic Chern bands assuming a finite interaction range. 
We also note that the underlying Berry curvature endows this metallic state with unquantized anomalous Hall effect~\cite{Sundaram1999,Haldane2004}. 
This ``anomalous Hall metal''~\cite{crepel2023anomalous} has been predicted and observed in $t$MoTe$_2$ at an extended filling range $\frac{2}{3}<\nu<1$~\cite{reddy2023toward,Anderson2024,anderson2025magnetoelectriccontrolhelicallight,park2025observationhightemperaturedissipationlessfractional}.

Our flat band model clearly displays a Fermi-liquid-like behavior at various filling factors in the vicinity of $\nu = 1$, as evidenced by the momentum occupation in the ED ground state $n(\k) = \langle c^{\dagger}_{\k} c_{\k} \rangle$ shown in Figs. \ref{fig:dispersion}(b)-(d).  
Remarkably, as $\mathcal K$ gets larger, the Fermi-liquid-like region expands down to $\nu = \frac{2}{3}$ and even lower fillings \cite{supplementary}, resulting in the destruction of the $\nu = \frac{2}{3}$ FQAH state shown in Fig. \ref{fig:instabilityFAQH}(c) and (d). We further calculate the ratio of kinetic energy to total energy of holes for the interacting ground state  at $\mathcal K=0.8$, $\eta \equiv \langle T \rangle/E_{\rm tot}$ where $\langle T \rangle = \sum_{\k} \epsilon(\k) (1-n(\k))$ \eqref{Hamiltonian_PH}. We find $\eta \approx 0.8$ for $\nu=\frac{2}{3}$, confirming the weakly interacting nature of holes. Moreover, $\langle T \rangle$ compares well with the kinetic energy $\langle T \rangle_0$ of the noninteracting hole gas with dispersion $\epsilon(\k)$, $\langle T \rangle_0 / \langle T \rangle \approx 0.92$. 
These ED results confirm the emergence of a hole Fermi liquid, {arising entirely from interaction effects without electron kinetic energy}, which provides a good starting point for studying many-body phases in flat Chern bands with nonuniform band geometry.  

Fermi liquids can develop pairing instabilities and become superconducting at low temperature. 
This motivates us to explore possible superconductivity in partially filled Chern bands, using a combination of analytical theory  and many-body numerics. 

{\it Pairing instability --} At low hole doping $\delta$, the minimum of $\epsilon(\bk)$ at the $\gamma$ point (Fig.~\ref{fig:dispersion}(a)) leads to a small Fermi surface defined by $\pi k_F^2 = \delta |\b_1 \times \b_2|$. 
Our system at low energy is a dilute hole gas with effective mass (we set $\hbar=1$ throughout this Letter) defined by
\begin{equation}\label{effective_mass}
    \frac{1}{m}=-\nabla^2_{\k}\Tr g(\bk)\Big|_{\gamma}\frac{\sum_{\bq\in\rm BZ} |\q|^2 V(\q)}{4A},
\end{equation}
and subject to residual interactions:
\begin{equation}\label{low_energy_model}\begin{split}
    \mathcal H_{\rm int}=\sum_{\bq\k\k'} \frac{V(\q)}{2A} \Lambda^*_{\k,\k+\q}\Lambda^*_{\k',\k'-\q}d^\dagger_{\k} d^\dagger_{\k'} d_{\k'-\q}d_{\k+\q}, 
\end{split}\end{equation}
where the form factors can be approximated at small momentum as
\begin{equation}
\Lambda_{\k,\k'}\approx1+i\ell^2(\k\times\k')/2-\ell^2(\k-\k')^2/4.    
\end{equation}
Here, the length scale $\ell$ is defined by $\ell^2 = \Tr g(\bk = 0)$, set by the quantum metric at the bottom of the hole dispersion. $1/\ell$ characterizes the extent of Bloch wave function variation in momentum space around $\gamma$. {Note that the nonuniformity parameter $\mathcal K$ controls the hole effective mass $m$.}

The pair scattering amplitude from an initial state $\ket{\k,-\k}$ to a final one $\ket{\k',-\k'}$ is the sum of a direct process with momentum transfer $\k - \k'$ and an exchange process with $\k + \k'$:

\begin{figure}
    \centering
    \includegraphics[width=0.9\linewidth]{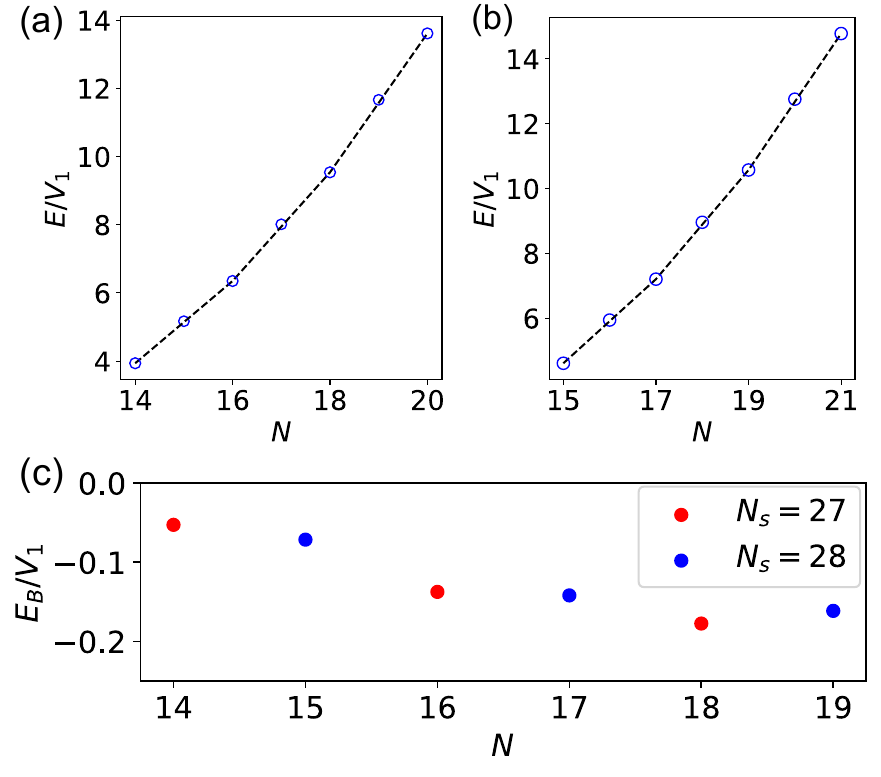}
    \caption{(a)-(b) Total energy $E$ at $\mathcal K = 0.8$ as a function of the number of particles $N$ for $N_s = 27$ (a) and $N_s = 28$ (b) where $N_s$ denote the number of sites of the finite-size system. (c) Binding energy $E_b = E(N+2) + E(N) - 2 E(N+1)$ evaluated at $\mathcal K = 0.8$. }
    \label{fig:bindingvsN}
\end{figure}

\begin{eqnarray}\label{antisymmetrized_ppinteraction}
\Gamma_{\k,\k'} &= & \frac{V(\k-\k')\Lambda_{\k',\k}\Lambda_{-\k',-\k}  -V(\k+\k')\Lambda_{-\k',\k}\Lambda_{\k',-\k}}{2} \nonumber\\
&= &\; v_1  k_+ k'_- ( 1-3\ell^2(k^2+k'^2)/2) \\
&  & + v_1  k_- k'_+ (1-\ell^2(k^2+k'^2)/2) \nonumber\\
& &  + v_1\ell^4 k_+^3 k'^3_-/2 , \nonumber
\end{eqnarray}
% \end{widetext}
where $k_\pm=k_x\pm ik_y$ and we employed $\Lambda^*_{\k,\k'}=\Lambda_{\k',\k}$. 
The influence of quantum geometry on the pair scattering amplitude~\eqref{antisymmetrized_ppinteraction} is controlled by the dimensionless parameter $\ell k_F$.
To zero-th order in $\ell k_F$ ($\Lambda_{\k'\k}=1$), particle-particle interaction is repulsive in the $p$-wave channel and of equal strength for relative angular momentum $L^z=\pm1$. 
However, the degeneracy between $L^z = \pm 1$ channels is lifted at the next leading order $(\ell  k_F)^2$. Moreover, a net repulsion at the order of $(\ell k_F)^4$ arises in the $L^z=-3$ channel  but not for $L^z=+3$~\footnote{Pairs with relative angular momentum $L^z = -(2m+1) < 0$ ($m>0$) experience repulsive interactions, with coupling strengths that are suppressed by a factor of $( \ell k_F)^{4m}$ with respect to the $V_1$-pseudopotential repulsion $v_1 k_F^2$}. 
The unequal particle-particle interaction strength at $\pm L$ reflects the chiral nature of the underlying Chern band wave functions.

While the bare interaction is nonattractive in any angular momentum channel, pairing can arise from the overscreening of the interaction between carriers.  
Within the random phase approximation and in the static limit, the screened interaction reads
\begin{equation}\label{screened}
\tilde V(\q)%=V(\bq)\varepsilon^{-1}(\q,0)
=\frac{V(\q)}{1+V(\q) \chi(\q,0)}.
\end{equation}
The screened particle-particle interaction $\tilde{\Gamma}_{\k,\k'}$ is obtained by replacing $V(\q)\to\tilde V(\q)$ in Eq.~\eqref{antisymmetrized_ppinteraction}.
We solve the linearized gap equation with $\tilde{\Gamma}_{\k,\k'}$~\cite{supplementary} and find that the leading superconducting instability occurs in the chiral $f$-wave channel $L_z=+3$.  

To understand heuristically pairing from the screened interaction~\eqref{screened}, we expand the geometric series~\eqref{screened} at small $\q$ %for $q \le 2k_F$ 
and find that the screening of the interaction yields a contribution $-v^2_1\q^4D(0)$ with $D(0)=m/(2\pi)$ the density of states, corresponding to an attractive potential $\propto -\nabla^4 \delta(\r)$.   
In ordinary 2D Fermi gas with trivial band geometry ($\ell=0$), this second-order interaction does not contribute to particle-particle scattering with relative angular momentum $L^z = \pm 3$, and pairing interaction only emerges at higher order ~\cite{chubukov_1993}.
In contrast, when the band geometry is nontrivial as in our case ($\ell\neq 0$),  a net attraction in the $L^z = +3$ channel already appears at second order, given by $-v_1^2D(0)\ell^2 (k_-^3 k'^3_+)$, while there is no bare repulsion in the $L^z = +3$ channel. 
Thus, band geometry in the anomalous Hall metal boosts superconductivity from repulsive interaction in our flat Chern band.
Related results on dispersive bands have been discussed recently~\cite{geier2024chiraltopologicalsuperconductivityisospin,Shavit_2025,jahin2025enhancedkohnluttingertopologicalsuperconductivity,xu2025chiral,maymann2025pairingmechanismdictatestopology,dong2025controllabletheorysuperconductivitystrong}.  
The characteristic pairing strength on the Fermi surface is given by the parameter $\lambda = v_1^2 D(0) \ell^2 k_F^6 \propto \delta^3$, highlighting a strong doping dependence.

The resulting superconducting state has a chiral $f$-wave order parameter, $\langle d^\dagger_{\k} d^\dagger_{-\k} \rangle \propto k^3_+$, and is fully gapped. Since our system is spin polarized and possesses a single Fermi surface enclosing $\gamma$, this chiral $f$-wave state (assuming weak pairing as implied in our analysis) is a topological superconductor characterized by Chern number $C=3/2$~\footnote{The Chern number $3/2$ implies three Majorana modes (rather than complex fermions), yielding a half-integer quantized thermal Hall response.}. As a hallmark of topology, this superconductor hosts $N=3$ branches of chiral Majorana edge modes at the boundary, and exhibits quantized thermal Hall conductance.

\begin{figure}[t!]
    \centering
    \includegraphics[width=0.9\linewidth]{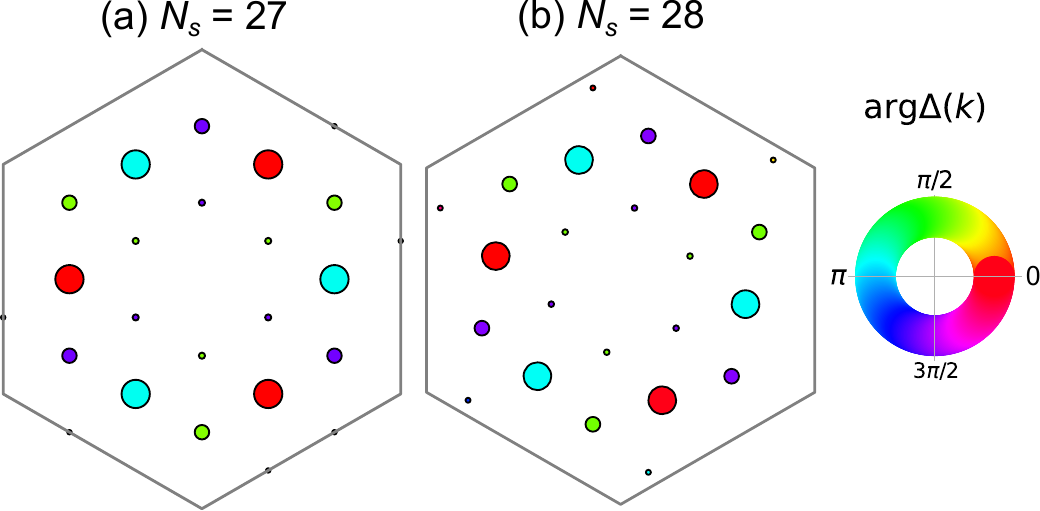}
    \caption{Gap function $\Delta(\bk) = \mel{\Psi_{N-2}}{c_{-\bk} c_{\bk}}{\Psi_{N}}$ for $\mathcal K$= 0.8 with $ N = 20 $ for $N_s = 27$ and $N = 21$ for $N_s = 28$. The size of the circle is proportional to $|\Delta(\k)|$ while the color denotes the phase.}
    \label{fig:pairwavefunction}
\end{figure}

{\it Numerical evidence of chiral $f$-wave pairing---} Motivated by our analytical results, which show that the anomalous Hall metal with short-range repulsion provides a favorable setting for superconducting instability, we now investigate our model using ED.
To this end, we analyze the ground state energy as a function of the particle number $N$ for systems with $N_s=27$ and $28$ unit cells under periodic boundary conditions (PBC), as shown in Fig.~\ref{fig:bindingvsN}(a) and (b), respectively.
In both cases, we find a finite pair binding energy $E_b \equiv E(N+2)+E(N)-2E(N+1)<0$ displayed in Fig.~\ref{fig:bindingvsN}(c), including at $2/3$ filling. In our convention, a negative $E_b$ means that the energy cost of adding two particles will be less than twice that of adding one particle, and therefore indicates pairing. {We also find that the energy of four doped carriers exceeds twice that of two, indicating the charge-$2e$ excitation as the lowest-energy charged excitation.}

Importantly, for system sizes $N_s=27$ and $28$, a negative binding energy is found only when the number of {\it holes} $N_h=N_s-N$ is odd. It may seem counterintuitive as the ground state of a superconductor normally has an even number of charges. This is however not the case here. A key feature of our {\it spinless} chiral superconductor in the weak pairing phase is that holes at opposite momenta $\k$ and $-\k$ are all paired up, except the one at $\k=0$ which does not have a partner to pair with. This leads to a reversed even-odd effect, i.e., the ground state of our system prefers to have an odd number of holes, exactly as observed by our ED calculation.

Notably, the binding energy constitutes a sizable fraction of the average kinetic energy per hole $E_K=\langle T\rangle/N_h$, which at $\mathcal K = 0.8$ and filling $2/3$ is found to be $E_K = 1.44 V_1$ by our ED calculation. In this case, the ratio $|E_b/2|/E_K \approx 0.06$ indicates sizable pairing, which is significantly larger than the value for most superconductors. 
Another important quantity for 2D superconductors is the superfluid stiffness $D_s$~\cite{emery1995importance,Levin2024RMP}. For systems with a parabolic dispersion, $D_s \sim \rho_h/m$ with $\rho_h=\delta /(|\a_1\times\a_2|)$ the density of carriers~\cite{Scalapino1992,Scalapino1993,Randeria2019,supplementary}.  
It is crucial to highlight that in a flat band system, the effective mass---and therefore the stiffness---arises entirely from the interaction effect and {\it nonuniform} band geometry, see Eq.~\eqref{effective_mass}.   In contrast, in the lowest Landau level limit ($\mathcal K=0$), the stiffness vanishes because charge-$2e$ Cooper pairs are immobile in a uniform magnetic field (see SM~\cite{supplementary}). Thus, a translationally invariant superconducting state cannot be formed in a Landau level. 
At $2/3$ filling and $\mathcal K=0.8$ we find the stiffness and binding energy are on the same order of magnitude:   $D_s/(|E_b|/2) \approx 5$. {In this regime, the superconducting $T_c$ is limited by the binding energy; for a typical bandwidth of $W\approx 8$meV, this gives $k_B T_c \sim |E_b|/2\approx 0.2~\mathrm{meV}$.}

Furthermore, to establish  the pairing symmetry we compute the superconducting order parameter $\Delta(\bk)\equiv\mel{\Psi_{N-2}}{c_{-\bk} c_{\bk}}{\Psi_{N}}$ with $\ket{\Psi_N}$ and $\ket{\Psi_{N-2}}$ many-body ground state in their respective particle number sectors \footnote{To calculate $\Delta(\k)$, we choose a gauge where the $\Delta(\k)$ is purely real at the momentum point where the absolute value $|\Delta(\k)|$ is maximum.}. 
Figure~\ref{fig:pairwavefunction}(a) and~\ref{fig:pairwavefunction}(b) displays the complex order parameter $\Delta(\bk)$ clearly revealing chiral $f$-wave symmetry. Notably, the order parameter is peaked around the Fermi surface, consistent with the weakly pairing regime. These ED results fully support a chiral $f$-wave topological superconductor, consistent with our theory.

{\it Discussion---} 
In this Letter, we have investigated the  interacting phase diagram of flat Chern bands tuned by inhomogeneous band geometry. We provided both numerical and analytical evidence for the emergence of chiral topological $f$-wave superconductivity around and {\it including} $\nu = 2/3$, which is driven by the fluctuation of quantum metric in $\k$ space.  
 
This superconducting phase can be understood from the pairing instability of a Fermi liquid parent state, which is an anomalous Hall metal with a singly connected Fermi surface. This is crucially distinct from recently proposed scenario of superconductivity of doped anyons in a fractional Chern insulator \cite{laughlinanyon,shi2024doping, divic2024anyon,shi2025anyondelocalizationtransitionsdisordered,nosov2025anyonsuperconductivityplateautransitions,pichler2025microscopicmechanismanyonsuperconductivity}.
Our theory differs from these proposals in fundamental ways. 
First, despite the nonperturbative nature of our theory, the superconducting state can be understood as a Fermi surface instability and is adiabatically connected to a chiral $f$-wave superconductor in the weak-pairing phase, {distinct from the topologically trivial BEC phase}. 
Second, in this topologically nontrivial regime, we find $N=3$ chiral Majorana edge modes—distinct from the hypothesized anyon superconductivity obtained by doping the $\nu=2/3$ FCI, which is argued to have an even number of edge modes~\cite{shi2024doping}.
{Thanks to its large binding energy, the superconducting state we found remains robust against perturbations, including bare band dispersion, longer-range interactions, and other variations to the flat Chern band model.}
% Importantly, our system belongs to the weak pairing (BCS) regime which is topologically nontrivial, distinct from the topologically trivial BEC phase. 

Possible signs of chiral superconductivity have been recently observed in twisted TMDs~\cite{xu2025signaturesunconventionalsuperconductivitynear}.  
Interestingly, superconductivity develops from a fairly good metal with $k_F\ell_{\rm mf}=2\pi/(\rho_{xx} e^2/h)\approx32$ that exhibits prominent anomalous Hall effect, consistent with the theoretical picture presented in this work. {Chiral $f$-wave topological superconductivity can be detected via the optical Kerr effect, half-integer quantized thermal Hall conductance, Majorana zero modes bound to vortex cores, and phase-sensitive probes of the order parameter.}

{\it Note added---} 
A subsequent work reports {evidence of} chiral $f$-wave superconductivity {and charge density wave} near a fractional Chern insulator in the lowest Landau level under a periodic potential~\cite{wang2025chiralsuperconductivitynearfractional}.

{\it Acknowledgments---} 
It is a pleasure to acknowledge useful discussions with Aidan Reddy. This material is based upon work supported by the Air Force Office of Scientific Research under award number FA2386-24-1-4043.
A. A. was supported by the Knut and Alice Wallenberg Foundation (KAW 2022.0348). L. F. was supported in part by a Simons Investigator Award from the Simons Foundation.

\bibliography{biblio}

\begin{thebibliography}{69}%
\makeatletter
\providecommand \@ifxundefined [1]{%
 \@ifx{#1\undefined}
}%
\providecommand \@ifnum [1]{%
 \ifnum #1\expandafter \@firstoftwo
 \else \expandafter \@secondoftwo
 \fi
}%
\providecommand \@ifx [1]{%
 \ifx #1\expandafter \@firstoftwo
 \else \expandafter \@secondoftwo
 \fi
}%
\providecommand \natexlab [1]{#1}%
\providecommand \enquote  [1]{``#1''}%
\providecommand \bibnamefont  [1]{#1}%
\providecommand \bibfnamefont [1]{#1}%
\providecommand \citenamefont [1]{#1}%
\providecommand \href@noop [0]{\@secondoftwo}%
\providecommand \href [0]{\begingroup \@sanitize@url \@href}%
\providecommand \@href[1]{\@@startlink{#1}\@@href}%
\providecommand \@@href[1]{\endgroup#1\@@endlink}%
\providecommand \@sanitize@url [0]{\catcode `\\12\catcode `\$12\catcode
  `\&12\catcode `\#12\catcode `\^12\catcode `\_12\catcode `\%12\relax}%
\providecommand \@@startlink[1]{}%
\providecommand \@@endlink[0]{}%
\providecommand \url  [0]{\begingroup\@sanitize@url \@url }%
\providecommand \@url [1]{\endgroup\@href {#1}{\urlprefix }}%
\providecommand \urlprefix  [0]{URL }%
\providecommand \Eprint [0]{\href }%
\providecommand \doibase [0]{https://doi.org/}%
\providecommand \selectlanguage [0]{\@gobble}%
\providecommand \bibinfo  [0]{\@secondoftwo}%
\providecommand \bibfield  [0]{\@secondoftwo}%
\providecommand \translation [1]{[#1]}%
\providecommand \BibitemOpen [0]{}%
\providecommand \bibitemStop [0]{}%
\providecommand \bibitemNoStop [0]{.\EOS\space}%
\providecommand \EOS [0]{\spacefactor3000\relax}%
\providecommand \BibitemShut  [1]{\csname bibitem#1\endcsname}%
\let\auto@bib@innerbib\@empty
%</preamble>
\bibitem [{\citenamefont {Cai}\ \emph {et~al.}(2023)\citenamefont {Cai},
  \citenamefont {Anderson}, \citenamefont {Wang}, \citenamefont {Zhang},
  \citenamefont {Liu}, \citenamefont {Holtzmann}, \citenamefont {Zhang},
  \citenamefont {Fan}, \citenamefont {Taniguchi}, \citenamefont {Watanabe},
  \citenamefont {Ran}, \citenamefont {Cao}, \citenamefont {Fu}, \citenamefont
  {Xiao}, \citenamefont {Yao},\ and\ \citenamefont {Xu}}]{Cai2023}%
  \BibitemOpen
  \bibfield  {author} {\bibinfo {author} {\bibfnamefont {J.}~\bibnamefont
  {Cai}}, \bibinfo {author} {\bibfnamefont {E.}~\bibnamefont {Anderson}},
  \bibinfo {author} {\bibfnamefont {C.}~\bibnamefont {Wang}}, \bibinfo {author}
  {\bibfnamefont {X.}~\bibnamefont {Zhang}}, \bibinfo {author} {\bibfnamefont
  {X.}~\bibnamefont {Liu}}, \bibinfo {author} {\bibfnamefont {W.}~\bibnamefont
  {Holtzmann}}, \bibinfo {author} {\bibfnamefont {Y.}~\bibnamefont {Zhang}},
  \bibinfo {author} {\bibfnamefont {F.}~\bibnamefont {Fan}}, \bibinfo {author}
  {\bibfnamefont {T.}~\bibnamefont {Taniguchi}}, \bibinfo {author}
  {\bibfnamefont {K.}~\bibnamefont {Watanabe}}, \bibinfo {author}
  {\bibfnamefont {Y.}~\bibnamefont {Ran}}, \bibinfo {author} {\bibfnamefont
  {T.}~\bibnamefont {Cao}}, \bibinfo {author} {\bibfnamefont {L.}~\bibnamefont
  {Fu}}, \bibinfo {author} {\bibfnamefont {D.}~\bibnamefont {Xiao}}, \bibinfo
  {author} {\bibfnamefont {W.}~\bibnamefont {Yao}},\ and\ \bibinfo {author}
  {\bibfnamefont {X.}~\bibnamefont {Xu}},\ }\bibfield  {title} {\bibinfo
  {title} {Signatures of fractional quantum anomalous hall states in twisted
  mote2},\ }\href {https://doi.org/10.1038/s41586-023-06289-w} {\bibfield
  {journal} {\bibinfo  {journal} {Nature}\ }\textbf {\bibinfo {volume} {622}},\
  \bibinfo {pages} {63} (\bibinfo {year} {2023})}\BibitemShut {NoStop}%
\bibitem [{\citenamefont {Zeng}\ \emph {et~al.}(2023)\citenamefont {Zeng},
  \citenamefont {Xia}, \citenamefont {Kang}, \citenamefont {Zhu}, \citenamefont
  {Kn{\"u}ppel}, \citenamefont {Vaswani}, \citenamefont {Watanabe},
  \citenamefont {Taniguchi}, \citenamefont {Mak},\ and\ \citenamefont
  {Shan}}]{zeng2023thermodynamic}%
  \BibitemOpen
  \bibfield  {author} {\bibinfo {author} {\bibfnamefont {Y.}~\bibnamefont
  {Zeng}}, \bibinfo {author} {\bibfnamefont {Z.}~\bibnamefont {Xia}}, \bibinfo
  {author} {\bibfnamefont {K.}~\bibnamefont {Kang}}, \bibinfo {author}
  {\bibfnamefont {J.}~\bibnamefont {Zhu}}, \bibinfo {author} {\bibfnamefont
  {P.}~\bibnamefont {Kn{\"u}ppel}}, \bibinfo {author} {\bibfnamefont
  {C.}~\bibnamefont {Vaswani}}, \bibinfo {author} {\bibfnamefont
  {K.}~\bibnamefont {Watanabe}}, \bibinfo {author} {\bibfnamefont
  {T.}~\bibnamefont {Taniguchi}}, \bibinfo {author} {\bibfnamefont {K.~F.}\
  \bibnamefont {Mak}},\ and\ \bibinfo {author} {\bibfnamefont {J.}~\bibnamefont
  {Shan}},\ }\bibfield  {title} {\bibinfo {title} {Thermodynamic evidence of
  fractional chern insulator in moir{\'e} mote2},\ }\href@noop {} {\bibfield
  {journal} {\bibinfo  {journal} {Nature}\ }\textbf {\bibinfo {volume} {622}},\
  \bibinfo {pages} {69} (\bibinfo {year} {2023})}\BibitemShut {NoStop}%
\bibitem [{\citenamefont {Park}\ \emph {et~al.}(2023)\citenamefont {Park},
  \citenamefont {Cai}, \citenamefont {Anderson}, \citenamefont {Zhang},
  \citenamefont {Zhu}, \citenamefont {Liu}, \citenamefont {Wang}, \citenamefont
  {Holtzmann}, \citenamefont {Hu}, \citenamefont {Liu}, \citenamefont
  {Taniguchi}, \citenamefont {Watanabe}, \citenamefont {Chu}, \citenamefont
  {Cao}, \citenamefont {Fu}, \citenamefont {Yao}, \citenamefont {Chang},
  \citenamefont {Cobden}, \citenamefont {Xiao},\ and\ \citenamefont
  {Xu}}]{xuParkObservationFractionallyQuantized2023}%
  \BibitemOpen
  \bibfield  {author} {\bibinfo {author} {\bibfnamefont {H.}~\bibnamefont
  {Park}}, \bibinfo {author} {\bibfnamefont {J.}~\bibnamefont {Cai}}, \bibinfo
  {author} {\bibfnamefont {E.}~\bibnamefont {Anderson}}, \bibinfo {author}
  {\bibfnamefont {Y.}~\bibnamefont {Zhang}}, \bibinfo {author} {\bibfnamefont
  {J.}~\bibnamefont {Zhu}}, \bibinfo {author} {\bibfnamefont {X.}~\bibnamefont
  {Liu}}, \bibinfo {author} {\bibfnamefont {C.}~\bibnamefont {Wang}}, \bibinfo
  {author} {\bibfnamefont {W.}~\bibnamefont {Holtzmann}}, \bibinfo {author}
  {\bibfnamefont {C.}~\bibnamefont {Hu}}, \bibinfo {author} {\bibfnamefont
  {Z.}~\bibnamefont {Liu}}, \bibinfo {author} {\bibfnamefont {T.}~\bibnamefont
  {Taniguchi}}, \bibinfo {author} {\bibfnamefont {K.}~\bibnamefont {Watanabe}},
  \bibinfo {author} {\bibfnamefont {J.-H.}\ \bibnamefont {Chu}}, \bibinfo
  {author} {\bibfnamefont {T.}~\bibnamefont {Cao}}, \bibinfo {author}
  {\bibfnamefont {L.}~\bibnamefont {Fu}}, \bibinfo {author} {\bibfnamefont
  {W.}~\bibnamefont {Yao}}, \bibinfo {author} {\bibfnamefont {C.-Z.}\
  \bibnamefont {Chang}}, \bibinfo {author} {\bibfnamefont {D.}~\bibnamefont
  {Cobden}}, \bibinfo {author} {\bibfnamefont {D.}~\bibnamefont {Xiao}},\ and\
  \bibinfo {author} {\bibfnamefont {X.}~\bibnamefont {Xu}},\ }\bibfield
  {title} {\bibinfo {title} {Observation of fractionally quantized anomalous
  hall effect},\ }\href {https://doi.org/10.1038/s41586-023-06536-0} {\bibfield
   {journal} {\bibinfo  {journal} {Nature}\ }\textbf {\bibinfo {volume}
  {622}},\ \bibinfo {pages} {74} (\bibinfo {year} {2023})}\BibitemShut
  {NoStop}%
\bibitem [{\citenamefont {Xu}\ \emph {et~al.}(2023)\citenamefont {Xu},
  \citenamefont {Sun}, \citenamefont {Jia}, \citenamefont {Liu}, \citenamefont
  {Xu}, \citenamefont {Li}, \citenamefont {Gu}, \citenamefont {Watanabe},
  \citenamefont {Taniguchi}, \citenamefont {Tong}, \citenamefont {Jia},
  \citenamefont {Shi}, \citenamefont {Jiang}, \citenamefont {Zhang},
  \citenamefont {Liu},\ and\ \citenamefont {Li}}]{PhysRevX.13.031037}%
  \BibitemOpen
  \bibfield  {author} {\bibinfo {author} {\bibfnamefont {F.}~\bibnamefont
  {Xu}}, \bibinfo {author} {\bibfnamefont {Z.}~\bibnamefont {Sun}}, \bibinfo
  {author} {\bibfnamefont {T.}~\bibnamefont {Jia}}, \bibinfo {author}
  {\bibfnamefont {C.}~\bibnamefont {Liu}}, \bibinfo {author} {\bibfnamefont
  {C.}~\bibnamefont {Xu}}, \bibinfo {author} {\bibfnamefont {C.}~\bibnamefont
  {Li}}, \bibinfo {author} {\bibfnamefont {Y.}~\bibnamefont {Gu}}, \bibinfo
  {author} {\bibfnamefont {K.}~\bibnamefont {Watanabe}}, \bibinfo {author}
  {\bibfnamefont {T.}~\bibnamefont {Taniguchi}}, \bibinfo {author}
  {\bibfnamefont {B.}~\bibnamefont {Tong}}, \bibinfo {author} {\bibfnamefont
  {J.}~\bibnamefont {Jia}}, \bibinfo {author} {\bibfnamefont {Z.}~\bibnamefont
  {Shi}}, \bibinfo {author} {\bibfnamefont {S.}~\bibnamefont {Jiang}}, \bibinfo
  {author} {\bibfnamefont {Y.}~\bibnamefont {Zhang}}, \bibinfo {author}
  {\bibfnamefont {X.}~\bibnamefont {Liu}},\ and\ \bibinfo {author}
  {\bibfnamefont {T.}~\bibnamefont {Li}},\ }\bibfield  {title} {\bibinfo
  {title} {Observation of integer and fractional quantum anomalous hall effects
  in twisted bilayer ${\mathrm{mote}}_{2}$},\ }\href
  {https://doi.org/10.1103/PhysRevX.13.031037} {\bibfield  {journal} {\bibinfo
  {journal} {Phys. Rev. X}\ }\textbf {\bibinfo {volume} {13}},\ \bibinfo
  {pages} {031037} (\bibinfo {year} {2023})}\BibitemShut {NoStop}%
\bibitem [{\citenamefont {Lu}\ \emph {et~al.}(2024)\citenamefont {Lu},
  \citenamefont {Han}, \citenamefont {Yao}, \citenamefont {Reddy},
  \citenamefont {Yang}, \citenamefont {Seo}, \citenamefont {Watanabe},
  \citenamefont {Taniguchi}, \citenamefont {Fu},\ and\ \citenamefont
  {Ju}}]{Lu2024Feb}%
  \BibitemOpen
  \bibfield  {author} {\bibinfo {author} {\bibfnamefont {Z.}~\bibnamefont
  {Lu}}, \bibinfo {author} {\bibfnamefont {T.}~\bibnamefont {Han}}, \bibinfo
  {author} {\bibfnamefont {Y.}~\bibnamefont {Yao}}, \bibinfo {author}
  {\bibfnamefont {A.~P.}\ \bibnamefont {Reddy}}, \bibinfo {author}
  {\bibfnamefont {J.}~\bibnamefont {Yang}}, \bibinfo {author} {\bibfnamefont
  {J.}~\bibnamefont {Seo}}, \bibinfo {author} {\bibfnamefont {K.}~\bibnamefont
  {Watanabe}}, \bibinfo {author} {\bibfnamefont {T.}~\bibnamefont {Taniguchi}},
  \bibinfo {author} {\bibfnamefont {L.}~\bibnamefont {Fu}},\ and\ \bibinfo
  {author} {\bibfnamefont {L.}~\bibnamefont {Ju}},\ }\bibfield  {title}
  {\bibinfo {title} {{Fractional quantum anomalous Hall effect in multilayer
  graphene}},\ }\href {https://doi.org/10.1038/s41586-023-07010-7} {\bibfield
  {journal} {\bibinfo  {journal} {Nature}\ }\textbf {\bibinfo {volume} {626}},\
  \bibinfo {pages} {759} (\bibinfo {year} {2024})}\BibitemShut {NoStop}%
\bibitem [{\citenamefont {Wu}\ \emph {et~al.}(2019)\citenamefont {Wu},
  \citenamefont {Lovorn}, \citenamefont {Tutuc}, \citenamefont {Martin},\ and\
  \citenamefont {MacDonald}}]{FW_PRL_2019}%
  \BibitemOpen
  \bibfield  {author} {\bibinfo {author} {\bibfnamefont {F.}~\bibnamefont
  {Wu}}, \bibinfo {author} {\bibfnamefont {T.}~\bibnamefont {Lovorn}}, \bibinfo
  {author} {\bibfnamefont {E.}~\bibnamefont {Tutuc}}, \bibinfo {author}
  {\bibfnamefont {I.}~\bibnamefont {Martin}},\ and\ \bibinfo {author}
  {\bibfnamefont {A.~H.}\ \bibnamefont {MacDonald}},\ }\bibfield  {title}
  {\bibinfo {title} {Topological insulators in twisted transition metal
  dichalcogenide homobilayers},\ }\href
  {https://doi.org/10.1103/PhysRevLett.122.086402} {\bibfield  {journal}
  {\bibinfo  {journal} {Phys. Rev. Lett.}\ }\textbf {\bibinfo {volume} {122}},\
  \bibinfo {pages} {086402} (\bibinfo {year} {2019})}\BibitemShut {NoStop}%
\bibitem [{\citenamefont {Devakul}\ \emph {et~al.}(2021)\citenamefont
  {Devakul}, \citenamefont {Crépel}, \citenamefont {Zhang},\ and\
  \citenamefont {Fu}}]{Devakul_2021}%
  \BibitemOpen
  \bibfield  {author} {\bibinfo {author} {\bibfnamefont {T.}~\bibnamefont
  {Devakul}}, \bibinfo {author} {\bibfnamefont {V.}~\bibnamefont {Crépel}},
  \bibinfo {author} {\bibfnamefont {Y.}~\bibnamefont {Zhang}},\ and\ \bibinfo
  {author} {\bibfnamefont {L.}~\bibnamefont {Fu}},\ }\bibfield  {title}
  {\bibinfo {title} {Magic in twisted transition metal dichalcogenide
  bilayers},\ }\bibfield  {journal} {\bibinfo  {journal} {Nature
  Communications}\ }\textbf {\bibinfo {volume} {12}},\ \href
  {https://doi.org/10.1038/s41467-021-27042-9} {10.1038/s41467-021-27042-9}
  (\bibinfo {year} {2021})\BibitemShut {NoStop}%
\bibitem [{\citenamefont {Lian}\ \emph {et~al.}(2021)\citenamefont {Lian},
  \citenamefont {Song}, \citenamefont {Regnault}, \citenamefont {Efetov},
  \citenamefont {Yazdani},\ and\ \citenamefont {Bernevig}}]{lian2021twisted}%
  \BibitemOpen
  \bibfield  {author} {\bibinfo {author} {\bibfnamefont {B.}~\bibnamefont
  {Lian}}, \bibinfo {author} {\bibfnamefont {Z.-D.}\ \bibnamefont {Song}},
  \bibinfo {author} {\bibfnamefont {N.}~\bibnamefont {Regnault}}, \bibinfo
  {author} {\bibfnamefont {D.~K.}\ \bibnamefont {Efetov}}, \bibinfo {author}
  {\bibfnamefont {A.}~\bibnamefont {Yazdani}},\ and\ \bibinfo {author}
  {\bibfnamefont {B.~A.}\ \bibnamefont {Bernevig}},\ }\bibfield  {title}
  {\bibinfo {title} {Twisted bilayer graphene. iv. exact insulator ground
  states and phase diagram},\ }\href
  {https://doi.org/10.1103/PhysRevB.103.205414} {\bibfield  {journal} {\bibinfo
   {journal} {Phys. Rev. B}\ }\textbf {\bibinfo {volume} {103}},\ \bibinfo
  {pages} {205414} (\bibinfo {year} {2021})}\BibitemShut {NoStop}%
\bibitem [{\citenamefont {Bultinck}\ \emph
  {et~al.}(2020{\natexlab{a}})\citenamefont {Bultinck}, \citenamefont {Khalaf},
  \citenamefont {Liu}, \citenamefont {Chatterjee}, \citenamefont {Vishwanath},\
  and\ \citenamefont {Zaletel}}]{bultinck2020ground}%
  \BibitemOpen
  \bibfield  {author} {\bibinfo {author} {\bibfnamefont {N.}~\bibnamefont
  {Bultinck}}, \bibinfo {author} {\bibfnamefont {E.}~\bibnamefont {Khalaf}},
  \bibinfo {author} {\bibfnamefont {S.}~\bibnamefont {Liu}}, \bibinfo {author}
  {\bibfnamefont {S.}~\bibnamefont {Chatterjee}}, \bibinfo {author}
  {\bibfnamefont {A.}~\bibnamefont {Vishwanath}},\ and\ \bibinfo {author}
  {\bibfnamefont {M.~P.}\ \bibnamefont {Zaletel}},\ }\bibfield  {title}
  {\bibinfo {title} {Ground state and hidden symmetry of magic-angle graphene
  at even integer filling},\ }\href
  {https://doi.org/10.1103/PhysRevX.10.031034} {\bibfield  {journal} {\bibinfo
  {journal} {Phys. Rev. X}\ }\textbf {\bibinfo {volume} {10}},\ \bibinfo
  {pages} {031034} (\bibinfo {year} {2020}{\natexlab{a}})}\BibitemShut
  {NoStop}%
\bibitem [{\citenamefont {Bultinck}\ \emph
  {et~al.}(2020{\natexlab{b}})\citenamefont {Bultinck}, \citenamefont
  {Chatterjee},\ and\ \citenamefont {Zaletel}}]{bultinck2020mechanism}%
  \BibitemOpen
  \bibfield  {author} {\bibinfo {author} {\bibfnamefont {N.}~\bibnamefont
  {Bultinck}}, \bibinfo {author} {\bibfnamefont {S.}~\bibnamefont
  {Chatterjee}},\ and\ \bibinfo {author} {\bibfnamefont {M.~P.}\ \bibnamefont
  {Zaletel}},\ }\bibfield  {title} {\bibinfo {title} {Mechanism for anomalous
  hall ferromagnetism in twisted bilayer graphene},\ }\href
  {https://doi.org/10.1103/PhysRevLett.124.166601} {\bibfield  {journal}
  {\bibinfo  {journal} {Phys. Rev. Lett.}\ }\textbf {\bibinfo {volume} {124}},\
  \bibinfo {pages} {166601} (\bibinfo {year} {2020}{\natexlab{b}})}\BibitemShut
  {NoStop}%
\bibitem [{\citenamefont {Repellin}\ \emph {et~al.}(2020)\citenamefont
  {Repellin}, \citenamefont {Dong}, \citenamefont {Zhang},\ and\ \citenamefont
  {Senthil}}]{repellin_ferromagnetism_2020}%
  \BibitemOpen
  \bibfield  {author} {\bibinfo {author} {\bibfnamefont {C.}~\bibnamefont
  {Repellin}}, \bibinfo {author} {\bibfnamefont {Z.}~\bibnamefont {Dong}},
  \bibinfo {author} {\bibfnamefont {Y.-H.}\ \bibnamefont {Zhang}},\ and\
  \bibinfo {author} {\bibfnamefont {T.}~\bibnamefont {Senthil}},\ }\bibfield
  {title} {\bibinfo {title} {Ferromagnetism in {Narrow} {Bands} of
  {Moir}{\textbackslash}'e {Superlattices}},\ }\href
  {https://doi.org/10.1103/PhysRevLett.124.187601} {\bibfield  {journal}
  {\bibinfo  {journal} {Physical Review Letters}\ }\textbf {\bibinfo {volume}
  {124}},\ \bibinfo {pages} {187601} (\bibinfo {year} {2020})},\ \bibinfo
  {note} {publisher: American Physical Society}\BibitemShut {NoStop}%
\bibitem [{\citenamefont {Wu}\ and\ \citenamefont
  {Das~Sarma}(2020)}]{wu2020collective}%
  \BibitemOpen
  \bibfield  {author} {\bibinfo {author} {\bibfnamefont {F.}~\bibnamefont
  {Wu}}\ and\ \bibinfo {author} {\bibfnamefont {S.}~\bibnamefont {Das~Sarma}},\
  }\bibfield  {title} {\bibinfo {title} {Collective excitations of quantum
  anomalous hall ferromagnets in twisted bilayer graphene},\ }\href
  {https://doi.org/10.1103/PhysRevLett.124.046403} {\bibfield  {journal}
  {\bibinfo  {journal} {Phys. Rev. Lett.}\ }\textbf {\bibinfo {volume} {124}},\
  \bibinfo {pages} {046403} (\bibinfo {year} {2020})}\BibitemShut {NoStop}%
\bibitem [{\citenamefont {Cr{\'e}pel}\ and\ \citenamefont
  {Fu}(2023)}]{crepel2023anomalous}%
  \BibitemOpen
  \bibfield  {author} {\bibinfo {author} {\bibfnamefont {V.}~\bibnamefont
  {Cr{\'e}pel}}\ and\ \bibinfo {author} {\bibfnamefont {L.}~\bibnamefont
  {Fu}},\ }\bibfield  {title} {\bibinfo {title} {Anomalous hall metal and
  fractional chern insulator in twisted transition metal dichalcogenides},\
  }\href@noop {} {\bibfield  {journal} {\bibinfo  {journal} {Physical Review
  B}\ }\textbf {\bibinfo {volume} {107}},\ \bibinfo {pages} {L201109} (\bibinfo
  {year} {2023})}\BibitemShut {NoStop}%
\bibitem [{\citenamefont {Reddy}\ \emph {et~al.}(2023)\citenamefont {Reddy},
  \citenamefont {Alsallom}, \citenamefont {Zhang}, \citenamefont {Devakul},\
  and\ \citenamefont {Fu}}]{reddyFQAH}%
  \BibitemOpen
  \bibfield  {author} {\bibinfo {author} {\bibfnamefont {A.~P.}\ \bibnamefont
  {Reddy}}, \bibinfo {author} {\bibfnamefont {F.}~\bibnamefont {Alsallom}},
  \bibinfo {author} {\bibfnamefont {Y.}~\bibnamefont {Zhang}}, \bibinfo
  {author} {\bibfnamefont {T.}~\bibnamefont {Devakul}},\ and\ \bibinfo {author}
  {\bibfnamefont {L.}~\bibnamefont {Fu}},\ }\bibfield  {title} {\bibinfo
  {title} {Fractional quantum anomalous hall states in twisted bilayer
  ${\mathrm{mote}}_{2}$ and ${\mathrm{wse}}_{2}$},\ }\href
  {https://doi.org/10.1103/PhysRevB.108.085117} {\bibfield  {journal} {\bibinfo
   {journal} {Phys. Rev. B}\ }\textbf {\bibinfo {volume} {108}},\ \bibinfo
  {pages} {085117} (\bibinfo {year} {2023})}\BibitemShut {NoStop}%
\bibitem [{\citenamefont {Anderson}\ \emph {et~al.}(2023)\citenamefont
  {Anderson}, \citenamefont {Fan}, \citenamefont {Cai}, \citenamefont
  {Holtzmann}, \citenamefont {Taniguchi}, \citenamefont {Watanabe},
  \citenamefont {Xiao}, \citenamefont {Yao},\ and\ \citenamefont
  {Xu}}]{Anderson2023}%
  \BibitemOpen
  \bibfield  {author} {\bibinfo {author} {\bibfnamefont {E.}~\bibnamefont
  {Anderson}}, \bibinfo {author} {\bibfnamefont {F.-R.}\ \bibnamefont {Fan}},
  \bibinfo {author} {\bibfnamefont {J.}~\bibnamefont {Cai}}, \bibinfo {author}
  {\bibfnamefont {W.}~\bibnamefont {Holtzmann}}, \bibinfo {author}
  {\bibfnamefont {T.}~\bibnamefont {Taniguchi}}, \bibinfo {author}
  {\bibfnamefont {K.}~\bibnamefont {Watanabe}}, \bibinfo {author}
  {\bibfnamefont {D.}~\bibnamefont {Xiao}}, \bibinfo {author} {\bibfnamefont
  {W.}~\bibnamefont {Yao}},\ and\ \bibinfo {author} {\bibfnamefont
  {X.}~\bibnamefont {Xu}},\ }\bibfield  {title} {\bibinfo {title} {Programming
  correlated magnetic states with gate-controlled moiré geometry},\ }\href
  {https://doi.org/10.1126/science.adg4268} {\bibfield  {journal} {\bibinfo
  {journal} {Science}\ }\textbf {\bibinfo {volume} {381}},\ \bibinfo {pages}
  {325–330} (\bibinfo {year} {2023})}\BibitemShut {NoStop}%
\bibitem [{\citenamefont {Li}\ \emph {et~al.}(2021)\citenamefont {Li},
  \citenamefont {Kumar}, \citenamefont {Sun},\ and\ \citenamefont
  {Lin}}]{Li2021PRR}%
  \BibitemOpen
  \bibfield  {author} {\bibinfo {author} {\bibfnamefont {H.}~\bibnamefont
  {Li}}, \bibinfo {author} {\bibfnamefont {U.}~\bibnamefont {Kumar}}, \bibinfo
  {author} {\bibfnamefont {K.}~\bibnamefont {Sun}},\ and\ \bibinfo {author}
  {\bibfnamefont {S.-Z.}\ \bibnamefont {Lin}},\ }\bibfield  {title} {\bibinfo
  {title} {Spontaneous fractional chern insulators in transition metal
  dichalcogenide moir\'e superlattices},\ }\href
  {https://doi.org/10.1103/PhysRevResearch.3.L032070} {\bibfield  {journal}
  {\bibinfo  {journal} {Phys. Rev. Res.}\ }\textbf {\bibinfo {volume} {3}},\
  \bibinfo {pages} {L032070} (\bibinfo {year} {2021})}\BibitemShut {NoStop}%
\bibitem [{\citenamefont {Tarnopolsky}\ \emph {et~al.}(2019)\citenamefont
  {Tarnopolsky}, \citenamefont {Kruchkov},\ and\ \citenamefont
  {Vishwanath}}]{Tarnopolsky_2019}%
  \BibitemOpen
  \bibfield  {author} {\bibinfo {author} {\bibfnamefont {G.}~\bibnamefont
  {Tarnopolsky}}, \bibinfo {author} {\bibfnamefont {A.~J.}\ \bibnamefont
  {Kruchkov}},\ and\ \bibinfo {author} {\bibfnamefont {A.}~\bibnamefont
  {Vishwanath}},\ }\bibfield  {title} {\bibinfo {title} {Origin of magic angles
  in twisted bilayer graphene},\ }\href
  {https://doi.org/10.1103/PhysRevLett.122.106405} {\bibfield  {journal}
  {\bibinfo  {journal} {Phys. Rev. Lett.}\ }\textbf {\bibinfo {volume} {122}},\
  \bibinfo {pages} {106405} (\bibinfo {year} {2019})}\BibitemShut {NoStop}%
\bibitem [{\citenamefont {Morales-Dur\'an}\ \emph {et~al.}(2023)\citenamefont
  {Morales-Dur\'an}, \citenamefont {Wang}, \citenamefont {Schleder},
  \citenamefont {Angeli}, \citenamefont {Zhu}, \citenamefont {Kaxiras},
  \citenamefont {Repellin},\ and\ \citenamefont {Cano}}]{Nicolas2023_prr}%
  \BibitemOpen
  \bibfield  {author} {\bibinfo {author} {\bibfnamefont {N.}~\bibnamefont
  {Morales-Dur\'an}}, \bibinfo {author} {\bibfnamefont {J.}~\bibnamefont
  {Wang}}, \bibinfo {author} {\bibfnamefont {G.~R.}\ \bibnamefont {Schleder}},
  \bibinfo {author} {\bibfnamefont {M.}~\bibnamefont {Angeli}}, \bibinfo
  {author} {\bibfnamefont {Z.}~\bibnamefont {Zhu}}, \bibinfo {author}
  {\bibfnamefont {E.}~\bibnamefont {Kaxiras}}, \bibinfo {author} {\bibfnamefont
  {C.}~\bibnamefont {Repellin}},\ and\ \bibinfo {author} {\bibfnamefont
  {J.}~\bibnamefont {Cano}},\ }\bibfield  {title} {\bibinfo {title}
  {Pressure-enhanced fractional chern insulators along a magic line in moir\'e
  transition metal dichalcogenides},\ }\href
  {https://doi.org/10.1103/PhysRevResearch.5.L032022} {\bibfield  {journal}
  {\bibinfo  {journal} {Phys. Rev. Res.}\ }\textbf {\bibinfo {volume} {5}},\
  \bibinfo {pages} {L032022} (\bibinfo {year} {2023})}\BibitemShut {NoStop}%
\bibitem [{\citenamefont {Morales-Dur\'an}\ \emph {et~al.}(2024)\citenamefont
  {Morales-Dur\'an}, \citenamefont {Wei}, \citenamefont {Shi},\ and\
  \citenamefont {MacDonald}}]{AllanNicolas_2024}%
  \BibitemOpen
  \bibfield  {author} {\bibinfo {author} {\bibfnamefont {N.}~\bibnamefont
  {Morales-Dur\'an}}, \bibinfo {author} {\bibfnamefont {N.}~\bibnamefont
  {Wei}}, \bibinfo {author} {\bibfnamefont {J.}~\bibnamefont {Shi}},\ and\
  \bibinfo {author} {\bibfnamefont {A.~H.}\ \bibnamefont {MacDonald}},\
  }\bibfield  {title} {\bibinfo {title} {Magic angles and fractional chern
  insulators in twisted homobilayer transition metal dichalcogenides},\ }\href
  {https://doi.org/10.1103/PhysRevLett.132.096602} {\bibfield  {journal}
  {\bibinfo  {journal} {Phys. Rev. Lett.}\ }\textbf {\bibinfo {volume} {132}},\
  \bibinfo {pages} {096602} (\bibinfo {year} {2024})}\BibitemShut {NoStop}%
\bibitem [{\citenamefont {Abouelkomsan}\ \emph {et~al.}(2023)\citenamefont
  {Abouelkomsan}, \citenamefont {Yang},\ and\ \citenamefont
  {Bergholtz}}]{Abouelkomsan_2023}%
  \BibitemOpen
  \bibfield  {author} {\bibinfo {author} {\bibfnamefont {A.}~\bibnamefont
  {Abouelkomsan}}, \bibinfo {author} {\bibfnamefont {K.}~\bibnamefont {Yang}},\
  and\ \bibinfo {author} {\bibfnamefont {E.~J.}\ \bibnamefont {Bergholtz}},\
  }\bibfield  {title} {\bibinfo {title} {Quantum metric induced phases in
  moiré materials},\ }\bibfield  {journal} {\bibinfo  {journal} {Physical
  Review Research}\ }\textbf {\bibinfo {volume} {5}},\ \href
  {https://doi.org/10.1103/physrevresearch.5.l012015}
  {10.1103/physrevresearch.5.l012015} (\bibinfo {year} {2023})\BibitemShut
  {NoStop}%
\bibitem [{\citenamefont {Reddy}\ and\ \citenamefont
  {Fu}(2023)}]{reddy2023toward}%
  \BibitemOpen
  \bibfield  {author} {\bibinfo {author} {\bibfnamefont {A.~P.}\ \bibnamefont
  {Reddy}}\ and\ \bibinfo {author} {\bibfnamefont {L.}~\bibnamefont {Fu}},\
  }\bibfield  {title} {\bibinfo {title} {Toward a global phase diagram of the
  fractional quantum anomalous hall effect},\ }\href@noop {} {\bibfield
  {journal} {\bibinfo  {journal} {Physical Review B}\ }\textbf {\bibinfo
  {volume} {108}},\ \bibinfo {pages} {245159} (\bibinfo {year}
  {2023})}\BibitemShut {NoStop}%
\bibitem [{\citenamefont {Han}\ \emph {et~al.}(2025)\citenamefont {Han},
  \citenamefont {Lu}, \citenamefont {Hadjri}, \citenamefont {Shi},
  \citenamefont {Wu}, \citenamefont {Xu}, \citenamefont {Yao}, \citenamefont
  {Cotten}, \citenamefont {Sedeh}, \citenamefont {Weldeyesus}, \citenamefont
  {Yang}, \citenamefont {Seo}, \citenamefont {Ye}, \citenamefont {Zhou},
  \citenamefont {Liu}, \citenamefont {Shi}, \citenamefont {Hua}, \citenamefont
  {Watanabe}, \citenamefont {Taniguchi}, \citenamefont {Xiong}, \citenamefont
  {Zumbühl}, \citenamefont {Fu},\ and\ \citenamefont
  {Ju}}]{han2025signatureschiralsuperconductivityrhombohedral}%
  \BibitemOpen
  \bibfield  {author} {\bibinfo {author} {\bibfnamefont {T.}~\bibnamefont
  {Han}}, \bibinfo {author} {\bibfnamefont {Z.}~\bibnamefont {Lu}}, \bibinfo
  {author} {\bibfnamefont {Z.}~\bibnamefont {Hadjri}}, \bibinfo {author}
  {\bibfnamefont {L.}~\bibnamefont {Shi}}, \bibinfo {author} {\bibfnamefont
  {Z.}~\bibnamefont {Wu}}, \bibinfo {author} {\bibfnamefont {W.}~\bibnamefont
  {Xu}}, \bibinfo {author} {\bibfnamefont {Y.}~\bibnamefont {Yao}}, \bibinfo
  {author} {\bibfnamefont {A.~A.}\ \bibnamefont {Cotten}}, \bibinfo {author}
  {\bibfnamefont {O.~S.}\ \bibnamefont {Sedeh}}, \bibinfo {author}
  {\bibfnamefont {H.}~\bibnamefont {Weldeyesus}}, \bibinfo {author}
  {\bibfnamefont {J.}~\bibnamefont {Yang}}, \bibinfo {author} {\bibfnamefont
  {J.}~\bibnamefont {Seo}}, \bibinfo {author} {\bibfnamefont {S.}~\bibnamefont
  {Ye}}, \bibinfo {author} {\bibfnamefont {M.}~\bibnamefont {Zhou}}, \bibinfo
  {author} {\bibfnamefont {H.}~\bibnamefont {Liu}}, \bibinfo {author}
  {\bibfnamefont {G.}~\bibnamefont {Shi}}, \bibinfo {author} {\bibfnamefont
  {Z.}~\bibnamefont {Hua}}, \bibinfo {author} {\bibfnamefont {K.}~\bibnamefont
  {Watanabe}}, \bibinfo {author} {\bibfnamefont {T.}~\bibnamefont {Taniguchi}},
  \bibinfo {author} {\bibfnamefont {P.}~\bibnamefont {Xiong}}, \bibinfo
  {author} {\bibfnamefont {D.~M.}\ \bibnamefont {Zumbühl}}, \bibinfo {author}
  {\bibfnamefont {L.}~\bibnamefont {Fu}},\ and\ \bibinfo {author}
  {\bibfnamefont {L.}~\bibnamefont {Ju}},\ }\href
  {https://arxiv.org/abs/2408.15233} {\bibinfo {title} {Signatures of chiral
  superconductivity in rhombohedral graphene}} (\bibinfo {year} {2025}),\
  \Eprint {https://arxiv.org/abs/2408.15233} {arXiv:2408.15233
  [cond-mat.mes-hall]} \BibitemShut {NoStop}%
\bibitem [{\citenamefont {Xu}\ \emph {et~al.}(2025{\natexlab{a}})\citenamefont
  {Xu}, \citenamefont {Sun}, \citenamefont {Li}, \citenamefont {Zheng},
  \citenamefont {Xu}, \citenamefont {Gao}, \citenamefont {Jia}, \citenamefont
  {Watanabe}, \citenamefont {Taniguchi}, \citenamefont {Tong}, \citenamefont
  {Lu}, \citenamefont {Jia}, \citenamefont {Shi}, \citenamefont {Jiang},
  \citenamefont {Zhang}, \citenamefont {Zhang}, \citenamefont {Lei},
  \citenamefont {Liu},\ and\ \citenamefont
  {Li}}]{xu2025signaturesunconventionalsuperconductivitynear}%
  \BibitemOpen
  \bibfield  {author} {\bibinfo {author} {\bibfnamefont {F.}~\bibnamefont
  {Xu}}, \bibinfo {author} {\bibfnamefont {Z.}~\bibnamefont {Sun}}, \bibinfo
  {author} {\bibfnamefont {J.}~\bibnamefont {Li}}, \bibinfo {author}
  {\bibfnamefont {C.}~\bibnamefont {Zheng}}, \bibinfo {author} {\bibfnamefont
  {C.}~\bibnamefont {Xu}}, \bibinfo {author} {\bibfnamefont {J.}~\bibnamefont
  {Gao}}, \bibinfo {author} {\bibfnamefont {T.}~\bibnamefont {Jia}}, \bibinfo
  {author} {\bibfnamefont {K.}~\bibnamefont {Watanabe}}, \bibinfo {author}
  {\bibfnamefont {T.}~\bibnamefont {Taniguchi}}, \bibinfo {author}
  {\bibfnamefont {B.}~\bibnamefont {Tong}}, \bibinfo {author} {\bibfnamefont
  {L.}~\bibnamefont {Lu}}, \bibinfo {author} {\bibfnamefont {J.}~\bibnamefont
  {Jia}}, \bibinfo {author} {\bibfnamefont {Z.}~\bibnamefont {Shi}}, \bibinfo
  {author} {\bibfnamefont {S.}~\bibnamefont {Jiang}}, \bibinfo {author}
  {\bibfnamefont {Y.}~\bibnamefont {Zhang}}, \bibinfo {author} {\bibfnamefont
  {Y.}~\bibnamefont {Zhang}}, \bibinfo {author} {\bibfnamefont
  {S.}~\bibnamefont {Lei}}, \bibinfo {author} {\bibfnamefont {X.}~\bibnamefont
  {Liu}},\ and\ \bibinfo {author} {\bibfnamefont {T.}~\bibnamefont {Li}},\
  }\href {https://arxiv.org/abs/2504.06972} {\bibinfo {title} {Signatures of
  unconventional superconductivity near reentrant and fractional quantum
  anomalous hall insulators}} (\bibinfo {year} {2025}{\natexlab{a}}),\ \Eprint
  {https://arxiv.org/abs/2504.06972} {arXiv:2504.06972 [cond-mat.mes-hall]}
  \BibitemShut {NoStop}%
\bibitem [{\citenamefont {Vanderbilt}(2018)}]{Vanderbilt_2018}%
  \BibitemOpen
  \bibfield  {author} {\bibinfo {author} {\bibfnamefont {D.}~\bibnamefont
  {Vanderbilt}},\ }\href@noop {} {\bibinfo {title} {Berry phases in electronic
  structure theory: Electric polarization, orbital magnetization and
  topological insulators}} (\bibinfo {year} {2018})\BibitemShut {NoStop}%
\bibitem [{\citenamefont {Fukui}\ \emph {et~al.}(2005)\citenamefont {Fukui},
  \citenamefont {Hatsugai},\ and\ \citenamefont {Suzuki}}]{Fukui_2005}%
  \BibitemOpen
  \bibfield  {author} {\bibinfo {author} {\bibfnamefont {T.}~\bibnamefont
  {Fukui}}, \bibinfo {author} {\bibfnamefont {Y.}~\bibnamefont {Hatsugai}},\
  and\ \bibinfo {author} {\bibfnamefont {H.}~\bibnamefont {Suzuki}},\
  }\bibfield  {title} {\bibinfo {title} {Chern numbers in discretized brillouin
  zone: Efficient method of computing (spin) hall conductances},\ }\href
  {https://doi.org/10.1143/jpsj.74.1674} {\bibfield  {journal} {\bibinfo
  {journal} {Journal of the Physical Society of Japan}\ }\textbf {\bibinfo
  {volume} {74}},\ \bibinfo {pages} {1674–1677} (\bibinfo {year}
  {2005})}\BibitemShut {NoStop}%
\bibitem [{\citenamefont {Dubrovin}\ and\ \citenamefont
  {Novikov}(1980)}]{Dubrovin_1980}%
  \BibitemOpen
  \bibfield  {author} {\bibinfo {author} {\bibfnamefont {B.}~\bibnamefont
  {Dubrovin}}\ and\ \bibinfo {author} {\bibfnamefont {S.}~\bibnamefont
  {Novikov}},\ }\bibfield  {title} {\bibinfo {title} {Ground states of a
  two-dimensional electron in a periodic magnetic field},\ }\href@noop {}
  {\bibfield  {journal} {\bibinfo  {journal} {Journal of Experimental and
  Theoretical Physics - J EXP THEOR PHYS}\ }\textbf {\bibinfo {volume} {52}}
  (\bibinfo {year} {1980})}\BibitemShut {NoStop}%
\bibitem [{\citenamefont {Aharonov}\ and\ \citenamefont
  {Casher}(1979)}]{Aharonov_1979}%
  \BibitemOpen
  \bibfield  {author} {\bibinfo {author} {\bibfnamefont {Y.}~\bibnamefont
  {Aharonov}}\ and\ \bibinfo {author} {\bibfnamefont {A.}~\bibnamefont
  {Casher}},\ }\bibfield  {title} {\bibinfo {title} {Ground state of a
  spin-\textonehalf{} charged particle in a two-dimensional magnetic field},\
  }\href {https://doi.org/10.1103/PhysRevA.19.2461} {\bibfield  {journal}
  {\bibinfo  {journal} {Phys. Rev. A}\ }\textbf {\bibinfo {volume} {19}},\
  \bibinfo {pages} {2461} (\bibinfo {year} {1979})}\BibitemShut {NoStop}%
\bibitem [{\citenamefont {Dong}\ \emph {et~al.}(2022)\citenamefont {Dong},
  \citenamefont {Wang},\ and\ \citenamefont
  {Fu}}]{dong2022diracelectronperiodicmagnetic}%
  \BibitemOpen
  \bibfield  {author} {\bibinfo {author} {\bibfnamefont {J.}~\bibnamefont
  {Dong}}, \bibinfo {author} {\bibfnamefont {J.}~\bibnamefont {Wang}},\ and\
  \bibinfo {author} {\bibfnamefont {L.}~\bibnamefont {Fu}},\ }\href
  {https://arxiv.org/abs/2208.10516} {\bibinfo {title} {Dirac electron under
  periodic magnetic field: Platform for fractional chern insulator and
  generalized wigner crystal}} (\bibinfo {year} {2022}),\ \Eprint
  {https://arxiv.org/abs/2208.10516} {arXiv:2208.10516 [cond-mat.mes-hall]}
  \BibitemShut {NoStop}%
\bibitem [{\citenamefont {Wang}\ \emph {et~al.}(2021)\citenamefont {Wang},
  \citenamefont {Cano}, \citenamefont {Millis}, \citenamefont {Liu},\ and\
  \citenamefont {Yang}}]{wang2021}%
  \BibitemOpen
  \bibfield  {author} {\bibinfo {author} {\bibfnamefont {J.}~\bibnamefont
  {Wang}}, \bibinfo {author} {\bibfnamefont {J.}~\bibnamefont {Cano}}, \bibinfo
  {author} {\bibfnamefont {A.~J.}\ \bibnamefont {Millis}}, \bibinfo {author}
  {\bibfnamefont {Z.}~\bibnamefont {Liu}},\ and\ \bibinfo {author}
  {\bibfnamefont {B.}~\bibnamefont {Yang}},\ }\bibfield  {title} {\bibinfo
  {title} {Exact landau level description of geometry and interaction in a
  flatband},\ }\href {https://doi.org/10.1103/PhysRevLett.127.246403}
  {\bibfield  {journal} {\bibinfo  {journal} {Phys. Rev. Lett.}\ }\textbf
  {\bibinfo {volume} {127}},\ \bibinfo {pages} {246403} (\bibinfo {year}
  {2021})}\BibitemShut {NoStop}%
\bibitem [{\citenamefont {Ledwith}\ \emph {et~al.}(2020)\citenamefont
  {Ledwith}, \citenamefont {Tarnopolsky}, \citenamefont {Khalaf},\ and\
  \citenamefont {Vishwanath}}]{Ledwith2020}%
  \BibitemOpen
  \bibfield  {author} {\bibinfo {author} {\bibfnamefont {P.~J.}\ \bibnamefont
  {Ledwith}}, \bibinfo {author} {\bibfnamefont {G.}~\bibnamefont
  {Tarnopolsky}}, \bibinfo {author} {\bibfnamefont {E.}~\bibnamefont
  {Khalaf}},\ and\ \bibinfo {author} {\bibfnamefont {A.}~\bibnamefont
  {Vishwanath}},\ }\bibfield  {title} {\bibinfo {title} {Fractional chern
  insulator states in twisted bilayer graphene: An analytical approach},\
  }\href {https://doi.org/10.1103/PhysRevResearch.2.023237} {\bibfield
  {journal} {\bibinfo  {journal} {Phys. Rev. Res.}\ }\textbf {\bibinfo {volume}
  {2}},\ \bibinfo {pages} {023237} (\bibinfo {year} {2020})}\BibitemShut
  {NoStop}%
\bibitem [{\citenamefont {Ledwith}\ \emph {et~al.}(2023)\citenamefont
  {Ledwith}, \citenamefont {Vishwanath},\ and\ \citenamefont
  {Parker}}]{Ledwith2023}%
  \BibitemOpen
  \bibfield  {author} {\bibinfo {author} {\bibfnamefont {P.~J.}\ \bibnamefont
  {Ledwith}}, \bibinfo {author} {\bibfnamefont {A.}~\bibnamefont
  {Vishwanath}},\ and\ \bibinfo {author} {\bibfnamefont {D.~E.}\ \bibnamefont
  {Parker}},\ }\bibfield  {title} {\bibinfo {title} {Vortexability: A unifying
  criterion for ideal fractional chern insulators},\ }\href
  {https://doi.org/10.1103/PhysRevB.108.205144} {\bibfield  {journal} {\bibinfo
   {journal} {Phys. Rev. B}\ }\textbf {\bibinfo {volume} {108}},\ \bibinfo
  {pages} {205144} (\bibinfo {year} {2023})}\BibitemShut {NoStop}%
\bibitem [{\citenamefont {Guerci}\ \emph {et~al.}(2024)\citenamefont {Guerci},
  \citenamefont {Wang},\ and\ \citenamefont
  {Mora}}]{guerci2024layerskyrmionsidealchern}%
  \BibitemOpen
  \bibfield  {author} {\bibinfo {author} {\bibfnamefont {D.}~\bibnamefont
  {Guerci}}, \bibinfo {author} {\bibfnamefont {J.}~\bibnamefont {Wang}},\ and\
  \bibinfo {author} {\bibfnamefont {C.}~\bibnamefont {Mora}},\ }\href
  {https://arxiv.org/abs/2408.12652} {\bibinfo {title} {Layer skyrmions for
  ideal chern bands and twisted bilayer graphene}} (\bibinfo {year} {2024}),\
  \Eprint {https://arxiv.org/abs/2408.12652} {arXiv:2408.12652
  [cond-mat.mes-hall]} \BibitemShut {NoStop}%
\bibitem [{\citenamefont {Paul}\ \emph {et~al.}(2023)\citenamefont {Paul},
  \citenamefont {Zhang},\ and\ \citenamefont {Fu}}]{Paul_2023}%
  \BibitemOpen
  \bibfield  {author} {\bibinfo {author} {\bibfnamefont {N.}~\bibnamefont
  {Paul}}, \bibinfo {author} {\bibfnamefont {Y.}~\bibnamefont {Zhang}},\ and\
  \bibinfo {author} {\bibfnamefont {L.}~\bibnamefont {Fu}},\ }\bibfield
  {title} {\bibinfo {title} {Giant proximity exchange and flat chern band in 2d
  magnet-semiconductor heterostructures},\ }\bibfield  {journal} {\bibinfo
  {journal} {Science Advances}\ }\textbf {\bibinfo {volume} {9}},\ \href
  {https://doi.org/10.1126/sciadv.abn1401} {10.1126/sciadv.abn1401} (\bibinfo
  {year} {2023})\BibitemShut {NoStop}%
\bibitem [{\citenamefont {Shi}\ \emph {et~al.}(2024)\citenamefont {Shi},
  \citenamefont {Morales-Durán}, \citenamefont {Khalaf},\ and\ \citenamefont
  {MacDonald}}]{Shi_2024}%
  \BibitemOpen
  \bibfield  {author} {\bibinfo {author} {\bibfnamefont {J.}~\bibnamefont
  {Shi}}, \bibinfo {author} {\bibfnamefont {N.}~\bibnamefont {Morales-Durán}},
  \bibinfo {author} {\bibfnamefont {E.}~\bibnamefont {Khalaf}},\ and\ \bibinfo
  {author} {\bibfnamefont {A.~H.}\ \bibnamefont {MacDonald}},\ }\bibfield
  {title} {\bibinfo {title} {Adiabatic approximation and aharonov-casher bands
  in twisted homobilayer transition metal dichalcogenides},\ }\bibfield
  {journal} {\bibinfo  {journal} {Physical Review B}\ }\textbf {\bibinfo
  {volume} {110}},\ \href {https://doi.org/10.1103/physrevb.110.035130}
  {10.1103/physrevb.110.035130} (\bibinfo {year} {2024})\BibitemShut {NoStop}%
\bibitem [{\citenamefont {Li}\ and\ \citenamefont {Wu}(2025)}]{wu2025}%
  \BibitemOpen
  \bibfield  {author} {\bibinfo {author} {\bibfnamefont {B.}~\bibnamefont
  {Li}}\ and\ \bibinfo {author} {\bibfnamefont {F.}~\bibnamefont {Wu}},\
  }\bibfield  {title} {\bibinfo {title} {Variational mapping of chern bands to
  landau levels: Application to fractional chern insulators in twisted
  ${\mathrm{mote}}_{2}$},\ }\href {https://doi.org/10.1103/PhysRevB.111.125122}
  {\bibfield  {journal} {\bibinfo  {journal} {Phys. Rev. B}\ }\textbf {\bibinfo
  {volume} {111}},\ \bibinfo {pages} {125122} (\bibinfo {year}
  {2025})}\BibitemShut {NoStop}%
\bibitem [{\citenamefont {Onishi}\ \emph {et~al.}(2025)\citenamefont {Onishi},
  \citenamefont {Avdoshkin},\ and\ \citenamefont
  {Fu}}]{onishi2025geometricboundstructurefactor}%
  \BibitemOpen
  \bibfield  {author} {\bibinfo {author} {\bibfnamefont {Y.}~\bibnamefont
  {Onishi}}, \bibinfo {author} {\bibfnamefont {A.}~\bibnamefont {Avdoshkin}},\
  and\ \bibinfo {author} {\bibfnamefont {L.}~\bibnamefont {Fu}},\ }\href
  {https://arxiv.org/abs/2412.02656} {\bibinfo {title} {Geometric bound on
  structure factor}} (\bibinfo {year} {2025}),\ \Eprint
  {https://arxiv.org/abs/2412.02656} {arXiv:2412.02656 [cond-mat.mes-hall]}
  \BibitemShut {NoStop}%
\bibitem [{sup()}]{supplementary}%
  \BibitemOpen
  \href@noop {} {}\bibinfo {note} {See the Supplemental Material [URL] for
  details on the model, exact diagonalization simulations, and long-wavelength
  expansion.}\BibitemShut {Stop}%
\bibitem [{\citenamefont {Giuliani}\ and\ \citenamefont
  {Vignale}(2005)}]{Giuliani_Vignale_2005}%
  \BibitemOpen
  \bibfield  {author} {\bibinfo {author} {\bibfnamefont {G.}~\bibnamefont
  {Giuliani}}\ and\ \bibinfo {author} {\bibfnamefont {G.}~\bibnamefont
  {Vignale}},\ }\href@noop {} {\emph {\bibinfo {title} {Quantum Theory of the
  Electron Liquid}}}\ (\bibinfo  {publisher} {Cambridge University Press},\
  \bibinfo {year} {2005})\BibitemShut {NoStop}%
\bibitem [{\citenamefont {Ghazaryan}\ \emph {et~al.}(2021)\citenamefont
  {Ghazaryan}, \citenamefont {Holder}, \citenamefont {Serbyn},\ and\
  \citenamefont {Berg}}]{Ghazaryan_2021}%
  \BibitemOpen
  \bibfield  {author} {\bibinfo {author} {\bibfnamefont {A.}~\bibnamefont
  {Ghazaryan}}, \bibinfo {author} {\bibfnamefont {T.}~\bibnamefont {Holder}},
  \bibinfo {author} {\bibfnamefont {M.}~\bibnamefont {Serbyn}},\ and\ \bibinfo
  {author} {\bibfnamefont {E.}~\bibnamefont {Berg}},\ }\bibfield  {title}
  {\bibinfo {title} {Unconventional superconductivity in systems with annular
  fermi surfaces: Application to rhombohedral trilayer graphene},\ }\bibfield
  {journal} {\bibinfo  {journal} {Physical Review Letters}\ }\textbf {\bibinfo
  {volume} {127}},\ \href {https://doi.org/10.1103/physrevlett.127.247001}
  {10.1103/physrevlett.127.247001} (\bibinfo {year} {2021})\BibitemShut
  {NoStop}%
\bibitem [{\citenamefont {Läuchli}\ \emph {et~al.}(2013)\citenamefont
  {Läuchli}, \citenamefont {Liu}, \citenamefont {Bergholtz},\ and\
  \citenamefont {Moessner}}]{lauchli_hierarchy_2013}%
  \BibitemOpen
  \bibfield  {author} {\bibinfo {author} {\bibfnamefont {A.~M.}\ \bibnamefont
  {Läuchli}}, \bibinfo {author} {\bibfnamefont {Z.}~\bibnamefont {Liu}},
  \bibinfo {author} {\bibfnamefont {E.~J.}\ \bibnamefont {Bergholtz}},\ and\
  \bibinfo {author} {\bibfnamefont {R.}~\bibnamefont {Moessner}},\ }\bibfield
  {title} {\bibinfo {title} {Hierarchy of {Fractional} {Chern} {Insulators} and
  {Competing} {Compressible} {States}},\ }\href
  {https://doi.org/10.1103/PhysRevLett.111.126802} {\bibfield  {journal}
  {\bibinfo  {journal} {Physical Review Letters}\ }\textbf {\bibinfo {volume}
  {111}},\ \bibinfo {pages} {126802} (\bibinfo {year} {2013})},\ \bibinfo
  {note} {publisher: American Physical Society}\BibitemShut {NoStop}%
\bibitem [{\citenamefont {Abouelkomsan}\ \emph {et~al.}(2020)\citenamefont
  {Abouelkomsan}, \citenamefont {Liu},\ and\ \citenamefont
  {Bergholtz}}]{abouelkomsan_particle-hole_2020}%
  \BibitemOpen
  \bibfield  {author} {\bibinfo {author} {\bibfnamefont {A.}~\bibnamefont
  {Abouelkomsan}}, \bibinfo {author} {\bibfnamefont {Z.}~\bibnamefont {Liu}},\
  and\ \bibinfo {author} {\bibfnamefont {E.~J.}\ \bibnamefont {Bergholtz}},\
  }\bibfield  {title} {\bibinfo {title} {Particle-{Hole} {Duality}, {Emergent}
  {Fermi} {Liquids}, and {Fractional} {Chern} {Insulators} in
  {Moir}{\textbackslash}'e {Flatbands}},\ }\href
  {https://doi.org/10.1103/PhysRevLett.124.106803} {\bibfield  {journal}
  {\bibinfo  {journal} {Physical Review Letters}\ }\textbf {\bibinfo {volume}
  {124}},\ \bibinfo {pages} {106803} (\bibinfo {year} {2020})},\ \bibinfo
  {note} {publisher: American Physical Society}\BibitemShut {NoStop}%
\bibitem [{\citenamefont {Liu}\ \emph {et~al.}(2025)\citenamefont {Liu},
  \citenamefont {Yang}, \citenamefont {Abouelkomsan}, \citenamefont {Liu},\
  and\ \citenamefont {Bergholtz}}]{liu_broken_2024}%
  \BibitemOpen
  \bibfield  {author} {\bibinfo {author} {\bibfnamefont {H.}~\bibnamefont
  {Liu}}, \bibinfo {author} {\bibfnamefont {K.}~\bibnamefont {Yang}}, \bibinfo
  {author} {\bibfnamefont {A.}~\bibnamefont {Abouelkomsan}}, \bibinfo {author}
  {\bibfnamefont {Z.}~\bibnamefont {Liu}},\ and\ \bibinfo {author}
  {\bibfnamefont {E.~J.}\ \bibnamefont {Bergholtz}},\ }\bibfield  {title}
  {\bibinfo {title} {Broken symmetry in ideal chern bands},\ }\href
  {https://doi.org/10.1103/PhysRevB.111.L201105} {\bibfield  {journal}
  {\bibinfo  {journal} {Phys. Rev. B}\ }\textbf {\bibinfo {volume} {111}},\
  \bibinfo {pages} {L201105} (\bibinfo {year} {2025})}\BibitemShut {NoStop}%
\bibitem [{\citenamefont {Sundaram}\ and\ \citenamefont
  {Niu}(1999)}]{Sundaram1999}%
  \BibitemOpen
  \bibfield  {author} {\bibinfo {author} {\bibfnamefont {G.}~\bibnamefont
  {Sundaram}}\ and\ \bibinfo {author} {\bibfnamefont {Q.}~\bibnamefont {Niu}},\
  }\bibfield  {title} {\bibinfo {title} {Wave-packet dynamics in slowly
  perturbed crystals: Gradient corrections and berry-phase effects},\ }\href
  {https://doi.org/10.1103/PhysRevB.59.14915} {\bibfield  {journal} {\bibinfo
  {journal} {Phys. Rev. B}\ }\textbf {\bibinfo {volume} {59}},\ \bibinfo
  {pages} {14915} (\bibinfo {year} {1999})}\BibitemShut {NoStop}%
\bibitem [{\citenamefont {Haldane}(2004)}]{Haldane2004}%
  \BibitemOpen
  \bibfield  {author} {\bibinfo {author} {\bibfnamefont {F.~D.~M.}\
  \bibnamefont {Haldane}},\ }\bibfield  {title} {\bibinfo {title} {Berry
  curvature on the fermi surface: Anomalous hall effect as a topological
  fermi-liquid property},\ }\href
  {https://doi.org/10.1103/PhysRevLett.93.206602} {\bibfield  {journal}
  {\bibinfo  {journal} {Phys. Rev. Lett.}\ }\textbf {\bibinfo {volume} {93}},\
  \bibinfo {pages} {206602} (\bibinfo {year} {2004})}\BibitemShut {NoStop}%
\bibitem [{\citenamefont {Anderson}\ \emph {et~al.}(2024)\citenamefont
  {Anderson}, \citenamefont {Cai}, \citenamefont {Reddy}, \citenamefont {Park},
  \citenamefont {Holtzmann}, \citenamefont {Davis}, \citenamefont {Taniguchi},
  \citenamefont {Watanabe}, \citenamefont {Smolenski}, \citenamefont
  {Imamoglu}, \citenamefont {Cao}, \citenamefont {Xiao}, \citenamefont {Fu},
  \citenamefont {Yao},\ and\ \citenamefont {Xu}}]{Anderson2024}%
  \BibitemOpen
  \bibfield  {author} {\bibinfo {author} {\bibfnamefont {E.}~\bibnamefont
  {Anderson}}, \bibinfo {author} {\bibfnamefont {J.}~\bibnamefont {Cai}},
  \bibinfo {author} {\bibfnamefont {A.~P.}\ \bibnamefont {Reddy}}, \bibinfo
  {author} {\bibfnamefont {H.}~\bibnamefont {Park}}, \bibinfo {author}
  {\bibfnamefont {W.}~\bibnamefont {Holtzmann}}, \bibinfo {author}
  {\bibfnamefont {K.}~\bibnamefont {Davis}}, \bibinfo {author} {\bibfnamefont
  {T.}~\bibnamefont {Taniguchi}}, \bibinfo {author} {\bibfnamefont
  {K.}~\bibnamefont {Watanabe}}, \bibinfo {author} {\bibfnamefont
  {T.}~\bibnamefont {Smolenski}}, \bibinfo {author} {\bibfnamefont
  {A.}~\bibnamefont {Imamoglu}}, \bibinfo {author} {\bibfnamefont
  {T.}~\bibnamefont {Cao}}, \bibinfo {author} {\bibfnamefont {D.}~\bibnamefont
  {Xiao}}, \bibinfo {author} {\bibfnamefont {L.}~\bibnamefont {Fu}}, \bibinfo
  {author} {\bibfnamefont {W.}~\bibnamefont {Yao}},\ and\ \bibinfo {author}
  {\bibfnamefont {X.}~\bibnamefont {Xu}},\ }\bibfield  {title} {\bibinfo
  {title} {Trion sensing of a zero-field composite fermi liquid},\ }\href
  {https://doi.org/10.1038/s41586-024-08134-0} {\bibfield  {journal} {\bibinfo
  {journal} {Nature}\ }\textbf {\bibinfo {volume} {635}},\ \bibinfo {pages}
  {590} (\bibinfo {year} {2024})}\BibitemShut {NoStop}%
\bibitem [{\citenamefont {Anderson}\ \emph {et~al.}(2025)\citenamefont
  {Anderson}, \citenamefont {Park}, \citenamefont {Yang}, \citenamefont {Cai},
  \citenamefont {Taniguchi}, \citenamefont {Watanabe}, \citenamefont {Fu},
  \citenamefont {Cao}, \citenamefont {Xiao},\ and\ \citenamefont
  {Xu}}]{anderson2025magnetoelectriccontrolhelicallight}%
  \BibitemOpen
  \bibfield  {author} {\bibinfo {author} {\bibfnamefont {E.}~\bibnamefont
  {Anderson}}, \bibinfo {author} {\bibfnamefont {H.}~\bibnamefont {Park}},
  \bibinfo {author} {\bibfnamefont {K.}~\bibnamefont {Yang}}, \bibinfo {author}
  {\bibfnamefont {J.}~\bibnamefont {Cai}}, \bibinfo {author} {\bibfnamefont
  {T.}~\bibnamefont {Taniguchi}}, \bibinfo {author} {\bibfnamefont
  {K.}~\bibnamefont {Watanabe}}, \bibinfo {author} {\bibfnamefont
  {L.}~\bibnamefont {Fu}}, \bibinfo {author} {\bibfnamefont {T.}~\bibnamefont
  {Cao}}, \bibinfo {author} {\bibfnamefont {D.}~\bibnamefont {Xiao}},\ and\
  \bibinfo {author} {\bibfnamefont {X.}~\bibnamefont {Xu}},\ }\href
  {https://arxiv.org/abs/2503.02810} {\bibinfo {title} {Magnetoelectric control
  of helical light emission in a moir\'e chern magnet}} (\bibinfo {year}
  {2025}),\ \Eprint {https://arxiv.org/abs/2503.02810} {arXiv:2503.02810
  [cond-mat.mes-hall]} \BibitemShut {NoStop}%
\bibitem [{\citenamefont {Park}\ \emph {et~al.}(2025)\citenamefont {Park},
  \citenamefont {Li}, \citenamefont {Hu}, \citenamefont {Beach}, \citenamefont
  {Gonçalves}, \citenamefont {Mendez-Valderrama}, \citenamefont
  {Herzog-Arbeitman}, \citenamefont {Taniguchi}, \citenamefont {Watanabe},
  \citenamefont {Cobden}, \citenamefont {Fu}, \citenamefont {Bernevig},
  \citenamefont {Regnault}, \citenamefont {Chu}, \citenamefont {Xiao},\ and\
  \citenamefont
  {Xu}}]{park2025observationhightemperaturedissipationlessfractional}%
  \BibitemOpen
  \bibfield  {author} {\bibinfo {author} {\bibfnamefont {H.}~\bibnamefont
  {Park}}, \bibinfo {author} {\bibfnamefont {W.}~\bibnamefont {Li}}, \bibinfo
  {author} {\bibfnamefont {C.}~\bibnamefont {Hu}}, \bibinfo {author}
  {\bibfnamefont {C.}~\bibnamefont {Beach}}, \bibinfo {author} {\bibfnamefont
  {M.}~\bibnamefont {Gonçalves}}, \bibinfo {author} {\bibfnamefont {J.~F.}\
  \bibnamefont {Mendez-Valderrama}}, \bibinfo {author} {\bibfnamefont
  {J.}~\bibnamefont {Herzog-Arbeitman}}, \bibinfo {author} {\bibfnamefont
  {T.}~\bibnamefont {Taniguchi}}, \bibinfo {author} {\bibfnamefont
  {K.}~\bibnamefont {Watanabe}}, \bibinfo {author} {\bibfnamefont
  {D.}~\bibnamefont {Cobden}}, \bibinfo {author} {\bibfnamefont
  {L.}~\bibnamefont {Fu}}, \bibinfo {author} {\bibfnamefont {B.~A.}\
  \bibnamefont {Bernevig}}, \bibinfo {author} {\bibfnamefont {N.}~\bibnamefont
  {Regnault}}, \bibinfo {author} {\bibfnamefont {J.-H.}\ \bibnamefont {Chu}},
  \bibinfo {author} {\bibfnamefont {D.}~\bibnamefont {Xiao}},\ and\ \bibinfo
  {author} {\bibfnamefont {X.}~\bibnamefont {Xu}},\ }\href
  {https://arxiv.org/abs/2503.10989} {\bibinfo {title} {Observation of
  high-temperature dissipationless fractional chern insulator}} (\bibinfo
  {year} {2025}),\ \Eprint {https://arxiv.org/abs/2503.10989} {arXiv:2503.10989
  [cond-mat.mes-hall]} \BibitemShut {NoStop}%
\bibitem [{Note1()}]{Note1}%
  \BibitemOpen
  \bibinfo {note} {Pairs with relative angular momentum $L^z = -(2m+1) < 0$
  ($m>0$) experience repulsive interactions, with coupling strengths that are
  suppressed by a factor of $( \ell k_F)^{4m}$ with respect to the
  $V_1$-pseudopotential repulsion $v_1 k_F^2$}\BibitemShut {NoStop}%
\bibitem [{\citenamefont {Chubukov}(1993)}]{chubukov_1993}%
  \BibitemOpen
  \bibfield  {author} {\bibinfo {author} {\bibfnamefont {A.~V.}\ \bibnamefont
  {Chubukov}},\ }\bibfield  {title} {\bibinfo {title} {Kohn-luttinger effect
  and the instability of a two-dimensional repulsive fermi liquid at t=0},\
  }\href {https://doi.org/10.1103/PhysRevB.48.1097} {\bibfield  {journal}
  {\bibinfo  {journal} {Phys. Rev. B}\ }\textbf {\bibinfo {volume} {48}},\
  \bibinfo {pages} {1097} (\bibinfo {year} {1993})}\BibitemShut {NoStop}%
\bibitem [{\citenamefont {Geier}\ \emph {et~al.}(2024)\citenamefont {Geier},
  \citenamefont {Davydova},\ and\ \citenamefont
  {Fu}}]{geier2024chiraltopologicalsuperconductivityisospin}%
  \BibitemOpen
  \bibfield  {author} {\bibinfo {author} {\bibfnamefont {M.}~\bibnamefont
  {Geier}}, \bibinfo {author} {\bibfnamefont {M.}~\bibnamefont {Davydova}},\
  and\ \bibinfo {author} {\bibfnamefont {L.}~\bibnamefont {Fu}},\ }\href
  {https://arxiv.org/abs/2409.13829} {\bibinfo {title} {Chiral and topological
  superconductivity in isospin polarized multilayer graphene}} (\bibinfo {year}
  {2024}),\ \Eprint {https://arxiv.org/abs/2409.13829} {arXiv:2409.13829
  [cond-mat.supr-con]} \BibitemShut {NoStop}%
\bibitem [{\citenamefont {Shavit}\ and\ \citenamefont
  {Alicea}(2025)}]{Shavit_2025}%
  \BibitemOpen
  \bibfield  {author} {\bibinfo {author} {\bibfnamefont {G.}~\bibnamefont
  {Shavit}}\ and\ \bibinfo {author} {\bibfnamefont {J.}~\bibnamefont
  {Alicea}},\ }\bibfield  {title} {\bibinfo {title} {Quantum geometric
  kohn-luttinger superconductivity},\ }\bibfield  {journal} {\bibinfo
  {journal} {Physical Review Letters}\ }\textbf {\bibinfo {volume} {134}},\
  \href {https://doi.org/10.1103/physrevlett.134.176001}
  {10.1103/physrevlett.134.176001} (\bibinfo {year} {2025})\BibitemShut
  {NoStop}%
\bibitem [{\citenamefont {Jahin}\ and\ \citenamefont
  {Lin}(2025)}]{jahin2025enhancedkohnluttingertopologicalsuperconductivity}%
  \BibitemOpen
  \bibfield  {author} {\bibinfo {author} {\bibfnamefont {A.}~\bibnamefont
  {Jahin}}\ and\ \bibinfo {author} {\bibfnamefont {S.-Z.}\ \bibnamefont
  {Lin}},\ }\href {https://arxiv.org/abs/2411.09664} {\bibinfo {title}
  {Enhanced kohn-luttinger topological superconductivity in bands with
  nontrivial geometry}} (\bibinfo {year} {2025}),\ \Eprint
  {https://arxiv.org/abs/2411.09664} {arXiv:2411.09664 [cond-mat.supr-con]}
  \BibitemShut {NoStop}%
\bibitem [{\citenamefont {Xu}\ \emph {et~al.}(2025{\natexlab{b}})\citenamefont
  {Xu}, \citenamefont {Zou}, \citenamefont {Peshcherenko}, \citenamefont
  {Jahin}, \citenamefont {Li}, \citenamefont {Lin},\ and\ \citenamefont
  {Zhang}}]{xu2025chiral}%
  \BibitemOpen
  \bibfield  {author} {\bibinfo {author} {\bibfnamefont {C.}~\bibnamefont
  {Xu}}, \bibinfo {author} {\bibfnamefont {N.}~\bibnamefont {Zou}}, \bibinfo
  {author} {\bibfnamefont {N.}~\bibnamefont {Peshcherenko}}, \bibinfo {author}
  {\bibfnamefont {A.}~\bibnamefont {Jahin}}, \bibinfo {author} {\bibfnamefont
  {T.}~\bibnamefont {Li}}, \bibinfo {author} {\bibfnamefont {S.-Z.}\
  \bibnamefont {Lin}},\ and\ \bibinfo {author} {\bibfnamefont {Y.}~\bibnamefont
  {Zhang}},\ }\href@noop {} {\bibinfo {title} {Chiral superconductivity from
  spin polarized chern band in twisted mote$_2$}} (\bibinfo {year}
  {2025}{\natexlab{b}}),\ \Eprint {https://arxiv.org/abs/2504.07082}
  {arXiv:2504.07082 [cond-mat.supr-con]} \BibitemShut {NoStop}%
\bibitem [{\citenamefont {May-Mann}\ \emph {et~al.}(2025)\citenamefont
  {May-Mann}, \citenamefont {Helbig},\ and\ \citenamefont
  {Devakul}}]{maymann2025pairingmechanismdictatestopology}%
  \BibitemOpen
  \bibfield  {author} {\bibinfo {author} {\bibfnamefont {J.}~\bibnamefont
  {May-Mann}}, \bibinfo {author} {\bibfnamefont {T.}~\bibnamefont {Helbig}},\
  and\ \bibinfo {author} {\bibfnamefont {T.}~\bibnamefont {Devakul}},\ }\href
  {https://arxiv.org/abs/2503.05697} {\bibinfo {title} {How pairing mechanism
  dictates topology in valley-polarized superconductors with berry curvature}}
  (\bibinfo {year} {2025}),\ \Eprint {https://arxiv.org/abs/2503.05697}
  {arXiv:2503.05697 [cond-mat.supr-con]} \BibitemShut {NoStop}%
\bibitem [{\citenamefont {Dong}\ and\ \citenamefont
  {Lee}(2025)}]{dong2025controllabletheorysuperconductivitystrong}%
  \BibitemOpen
  \bibfield  {author} {\bibinfo {author} {\bibfnamefont {Z.}~\bibnamefont
  {Dong}}\ and\ \bibinfo {author} {\bibfnamefont {P.~A.}\ \bibnamefont {Lee}},\
  }\href {https://arxiv.org/abs/2503.11079} {\bibinfo {title} {A controllable
  theory of superconductivity due to strong repulsion in a polarized band}}
  (\bibinfo {year} {2025}),\ \Eprint {https://arxiv.org/abs/2503.11079}
  {arXiv:2503.11079 [cond-mat.supr-con]} \BibitemShut {NoStop}%
\bibitem [{Note2()}]{Note2}%
  \BibitemOpen
  \bibinfo {note} {The Chern number $3/2$ implies three Majorana modes (rather
  than complex fermions), yielding a half-integer quantized thermal Hall
  response.}\BibitemShut {Stop}%
\bibitem [{\citenamefont {Emery}\ and\ \citenamefont
  {Kivelson}(1995)}]{emery1995importance}%
  \BibitemOpen
  \bibfield  {author} {\bibinfo {author} {\bibfnamefont {V.}~\bibnamefont
  {Emery}}\ and\ \bibinfo {author} {\bibfnamefont {S.}~\bibnamefont
  {Kivelson}},\ }\bibfield  {title} {\bibinfo {title} {Importance of phase
  fluctuations in superconductors with small superfluid density},\ }\href@noop
  {} {\bibfield  {journal} {\bibinfo  {journal} {Nature}\ }\textbf {\bibinfo
  {volume} {374}},\ \bibinfo {pages} {434} (\bibinfo {year}
  {1995})}\BibitemShut {NoStop}%
\bibitem [{\citenamefont {Chen}\ \emph {et~al.}(2024)\citenamefont {Chen},
  \citenamefont {Wang}, \citenamefont {Boyack}, \citenamefont {Yang},\ and\
  \citenamefont {Levin}}]{Levin2024RMP}%
  \BibitemOpen
  \bibfield  {author} {\bibinfo {author} {\bibfnamefont {Q.}~\bibnamefont
  {Chen}}, \bibinfo {author} {\bibfnamefont {Z.}~\bibnamefont {Wang}}, \bibinfo
  {author} {\bibfnamefont {R.}~\bibnamefont {Boyack}}, \bibinfo {author}
  {\bibfnamefont {S.}~\bibnamefont {Yang}},\ and\ \bibinfo {author}
  {\bibfnamefont {K.}~\bibnamefont {Levin}},\ }\bibfield  {title} {\bibinfo
  {title} {When superconductivity crosses over: From bcs to bec},\ }\href
  {https://doi.org/10.1103/RevModPhys.96.025002} {\bibfield  {journal}
  {\bibinfo  {journal} {Rev. Mod. Phys.}\ }\textbf {\bibinfo {volume} {96}},\
  \bibinfo {pages} {025002} (\bibinfo {year} {2024})}\BibitemShut {NoStop}%
\bibitem [{\citenamefont {Scalapino}\ \emph {et~al.}(1992)\citenamefont
  {Scalapino}, \citenamefont {White},\ and\ \citenamefont
  {Zhang}}]{Scalapino1992}%
  \BibitemOpen
  \bibfield  {author} {\bibinfo {author} {\bibfnamefont {D.~J.}\ \bibnamefont
  {Scalapino}}, \bibinfo {author} {\bibfnamefont {S.~R.}\ \bibnamefont
  {White}},\ and\ \bibinfo {author} {\bibfnamefont {S.~C.}\ \bibnamefont
  {Zhang}},\ }\bibfield  {title} {\bibinfo {title} {Superfluid density and the
  drude weight of the hubbard model},\ }\href
  {https://doi.org/10.1103/PhysRevLett.68.2830} {\bibfield  {journal} {\bibinfo
   {journal} {Phys. Rev. Lett.}\ }\textbf {\bibinfo {volume} {68}},\ \bibinfo
  {pages} {2830} (\bibinfo {year} {1992})}\BibitemShut {NoStop}%
\bibitem [{\citenamefont {Scalapino}\ \emph {et~al.}(1993)\citenamefont
  {Scalapino}, \citenamefont {White},\ and\ \citenamefont
  {Zhang}}]{Scalapino1993}%
  \BibitemOpen
  \bibfield  {author} {\bibinfo {author} {\bibfnamefont {D.~J.}\ \bibnamefont
  {Scalapino}}, \bibinfo {author} {\bibfnamefont {S.~R.}\ \bibnamefont
  {White}},\ and\ \bibinfo {author} {\bibfnamefont {S.}~\bibnamefont {Zhang}},\
  }\bibfield  {title} {\bibinfo {title} {Insulator, metal, or superconductor:
  The criteria},\ }\href {https://doi.org/10.1103/PhysRevB.47.7995} {\bibfield
  {journal} {\bibinfo  {journal} {Phys. Rev. B}\ }\textbf {\bibinfo {volume}
  {47}},\ \bibinfo {pages} {7995} (\bibinfo {year} {1993})}\BibitemShut
  {NoStop}%
\bibitem [{\citenamefont {Hazra}\ \emph {et~al.}(2019)\citenamefont {Hazra},
  \citenamefont {Verma},\ and\ \citenamefont {Randeria}}]{Randeria2019}%
  \BibitemOpen
  \bibfield  {author} {\bibinfo {author} {\bibfnamefont {T.}~\bibnamefont
  {Hazra}}, \bibinfo {author} {\bibfnamefont {N.}~\bibnamefont {Verma}},\ and\
  \bibinfo {author} {\bibfnamefont {M.}~\bibnamefont {Randeria}},\ }\bibfield
  {title} {\bibinfo {title} {Bounds on the superconducting transition
  temperature: Applications to twisted bilayer graphene and cold atoms},\
  }\href {https://doi.org/10.1103/PhysRevX.9.031049} {\bibfield  {journal}
  {\bibinfo  {journal} {Phys. Rev. X}\ }\textbf {\bibinfo {volume} {9}},\
  \bibinfo {pages} {031049} (\bibinfo {year} {2019})}\BibitemShut {NoStop}%
\bibitem [{Note3()}]{Note3}%
  \BibitemOpen
  \bibinfo {note} {To calculate $\Delta ({\protect \boldsymbol {k}})$, we
  choose a gauge where the $\Delta ({\protect \boldsymbol {k}})$ is purely real
  at the momentum point where the absolute value $|\Delta ({\protect
  \boldsymbol {k}})|$ is maximum.}\BibitemShut {Stop}%
\bibitem [{\citenamefont {Laughlin}(1988)}]{laughlinanyon}%
  \BibitemOpen
  \bibfield  {author} {\bibinfo {author} {\bibfnamefont {R.~B.}\ \bibnamefont
  {Laughlin}},\ }\bibfield  {title} {\bibinfo {title} {Superconducting ground
  state of noninteracting particles obeying fractional statistics},\ }\href
  {https://doi.org/10.1103/PhysRevLett.60.2677} {\bibfield  {journal} {\bibinfo
   {journal} {Phys. Rev. Lett.}\ }\textbf {\bibinfo {volume} {60}},\ \bibinfo
  {pages} {2677} (\bibinfo {year} {1988})}\BibitemShut {NoStop}%
\bibitem [{\citenamefont {Shi}\ and\ \citenamefont
  {Senthil}(2024)}]{shi2024doping}%
  \BibitemOpen
  \bibfield  {author} {\bibinfo {author} {\bibfnamefont {Z.~D.}\ \bibnamefont
  {Shi}}\ and\ \bibinfo {author} {\bibfnamefont {T.}~\bibnamefont {Senthil}},\
  }\bibfield  {title} {\bibinfo {title} {Doping a fractional quantum anomalous
  hall insulator},\ }\href@noop {} {\bibfield  {journal} {\bibinfo  {journal}
  {arXiv preprint arXiv:2409.20567}\ } (\bibinfo {year} {2024})}\BibitemShut
  {NoStop}%
\bibitem [{\citenamefont {Divic}\ \emph {et~al.}(2024)\citenamefont {Divic},
  \citenamefont {Cr{\'e}pel}, \citenamefont {Soejima}, \citenamefont {Song},
  \citenamefont {Millis}, \citenamefont {Zaletel},\ and\ \citenamefont
  {Vishwanath}}]{divic2024anyon}%
  \BibitemOpen
  \bibfield  {author} {\bibinfo {author} {\bibfnamefont {S.}~\bibnamefont
  {Divic}}, \bibinfo {author} {\bibfnamefont {V.}~\bibnamefont {Cr{\'e}pel}},
  \bibinfo {author} {\bibfnamefont {T.}~\bibnamefont {Soejima}}, \bibinfo
  {author} {\bibfnamefont {X.-Y.}\ \bibnamefont {Song}}, \bibinfo {author}
  {\bibfnamefont {A.}~\bibnamefont {Millis}}, \bibinfo {author} {\bibfnamefont
  {M.~P.}\ \bibnamefont {Zaletel}},\ and\ \bibinfo {author} {\bibfnamefont
  {A.}~\bibnamefont {Vishwanath}},\ }\bibfield  {title} {\bibinfo {title}
  {Anyon superconductivity from topological criticality in a hofstadter-hubbard
  model},\ }\href@noop {} {\bibfield  {journal} {\bibinfo  {journal} {arXiv
  preprint arXiv:2410.18175}\ } (\bibinfo {year} {2024})}\BibitemShut {NoStop}%
\bibitem [{\citenamefont {Shi}\ and\ \citenamefont
  {Senthil}(2025)}]{shi2025anyondelocalizationtransitionsdisordered}%
  \BibitemOpen
  \bibfield  {author} {\bibinfo {author} {\bibfnamefont {Z.~D.}\ \bibnamefont
  {Shi}}\ and\ \bibinfo {author} {\bibfnamefont {T.}~\bibnamefont {Senthil}},\
  }\href {https://arxiv.org/abs/2506.02128} {\bibinfo {title} {Anyon
  delocalization transitions out of a disordered fqah insulator}} (\bibinfo
  {year} {2025}),\ \Eprint {https://arxiv.org/abs/2506.02128} {arXiv:2506.02128
  [cond-mat.str-el]} \BibitemShut {NoStop}%
\bibitem [{\citenamefont {Nosov}\ \emph {et~al.}(2025)\citenamefont {Nosov},
  \citenamefont {Han},\ and\ \citenamefont
  {Khalaf}}]{nosov2025anyonsuperconductivityplateautransitions}%
  \BibitemOpen
  \bibfield  {author} {\bibinfo {author} {\bibfnamefont {P.~A.}\ \bibnamefont
  {Nosov}}, \bibinfo {author} {\bibfnamefont {Z.}~\bibnamefont {Han}},\ and\
  \bibinfo {author} {\bibfnamefont {E.}~\bibnamefont {Khalaf}},\ }\href
  {https://arxiv.org/abs/2506.02108} {\bibinfo {title} {Anyon superconductivity
  and plateau transitions in doped fractional quantum anomalous hall
  insulators}} (\bibinfo {year} {2025}),\ \Eprint
  {https://arxiv.org/abs/2506.02108} {arXiv:2506.02108 [cond-mat.str-el]}
  \BibitemShut {NoStop}%
\bibitem [{\citenamefont {Pichler}\ \emph {et~al.}(2025)\citenamefont
  {Pichler}, \citenamefont {Kuhlenkamp}, \citenamefont {Knap},\ and\
  \citenamefont
  {Vishwanath}}]{pichler2025microscopicmechanismanyonsuperconductivity}%
  \BibitemOpen
  \bibfield  {author} {\bibinfo {author} {\bibfnamefont {F.}~\bibnamefont
  {Pichler}}, \bibinfo {author} {\bibfnamefont {C.}~\bibnamefont {Kuhlenkamp}},
  \bibinfo {author} {\bibfnamefont {M.}~\bibnamefont {Knap}},\ and\ \bibinfo
  {author} {\bibfnamefont {A.}~\bibnamefont {Vishwanath}},\ }\href
  {https://arxiv.org/abs/2506.08000} {\bibinfo {title} {Microscopic mechanism
  of anyon superconductivity emerging from fractional chern insulators}}
  (\bibinfo {year} {2025}),\ \Eprint {https://arxiv.org/abs/2506.08000}
  {arXiv:2506.08000 [cond-mat.str-el]} \BibitemShut {NoStop}%
\bibitem [{\citenamefont {Wang}\ and\ \citenamefont
  {Zaletel}(2025)}]{wang2025chiralsuperconductivitynearfractional}%
  \BibitemOpen
  \bibfield  {author} {\bibinfo {author} {\bibfnamefont {T.}~\bibnamefont
  {Wang}}\ and\ \bibinfo {author} {\bibfnamefont {M.~P.}\ \bibnamefont
  {Zaletel}},\ }\href {https://arxiv.org/abs/2507.07921} {\bibinfo {title}
  {Chiral superconductivity near a fractional chern insulator}} (\bibinfo
  {year} {2025}),\ \Eprint {https://arxiv.org/abs/2507.07921} {arXiv:2507.07921
  [cond-mat.str-el]} \BibitemShut {NoStop}%
\end{thebibliography}%

% ------------------------------------------------------------------------ %
% ------------------------------------------------------------------------ %
% ------------------------------------------------------------------------ %

\onecolumngrid
\newpage
\makeatletter

\begin{center}
\textbf{\large Supplementary materials for: ``\@title ''} \\[10pt]
Daniele Guerci$^{1,*}$, Ahmed Abouelkomsan$^{1,*}$ and Liang Fu$^1$ \\
\textit{$^1$Department of Physics, Massachusetts Institute of Technology, Cambridge, MA-02139, USA}\\
\end{center}
$^*$These authors contributed equally to this work.
\vspace{10pt}

\setcounter{page}{1} % Start from page 1 (you can use 0, but journals start at 1)
\setcounter{figure}{0}
\setcounter{section}{0}
\setcounter{equation}{0}

\renewcommand{\thefigure}{S\@arabic\c@figure}
\makeatother

\appendix

These supplementary materials contain the details of the flat Chern band model as well as additional calculations supporting the results presented in the main text.
Sec.~\ref{app_sec:model} contains details on the model Hamiltonian employed in our analysis.  
In Sec.~\ref{app_sec:ahm} we characterize the single hole interaction-induced dispersion relation. 
Sec.~\ref{app_sec:long_wavelength} provides a derivation of the long wavelength theory employed to derive the leading weak coupling superconducting instability.
Finally, in Sec.~\ref{app_sec:ed} we provide additional ED results in support to our theory.

\section{Flat Chern band model}\label{app_sec:model}

The Hamiltonian is given by the interaction only: 
\begin{equation}\label{app:hamiltonian}
     H_{\rm int}=\frac{1}{2A}\sum_{\k_1\cdots \k_4}H_{\k_1,\k_2;\k_3,\k_4}\, c^\dagger_{\k_1 }c^\dagger_{\k_2}c_{\k_3}c_{\k_4},
\end{equation}
where we have introduced $A=|\bm L_1\times \bm L_2|$ with $\bm L_{1/2}$ defining sides of the torus. 
The interaction matrix element reads: 
\begin{equation}\label{matrix_element}
    H_{\k_1,\k_2;\k_3,\k_4}= \sum_{\G} \delta_{\k_1+\k_2-\k_3-\k_4,\Delta \G}\, V_{\k_1-\k_4-\G}\Lambda_{\k_1,\k_4+\G}\,\Lambda_{\bk_2,\bk_3+\Delta \G-\G},
\end{equation}
where $\G$ and $\Delta\G$ are reciprocal lattice vectors, and
\begin{equation}
    V_{\q}=\int d^2\br e^{-i\bq\cdot \br} V(\br), 
\end{equation}
with $V(\br)$ real-space interaction. 
We will consider in our work short-range interaction.
Finally, we have introduced the form factors defined as:
\begin{equation}
    \Lambda_{\k,\p+\G}=\int_{\rm UC}\frac{d^2r}{\Omega}e^{-i\G\cdot\r}u^*_{\bk }(\bm r) u_{\bp}(\bm r),%=\mel{u_{\bk}}{V^{\bm g}}{u_{\bm p}},
\end{equation}
where $\Omega=|\bm a_1\times \bm a_2|$ and $u_{\bk}(\br)$ the cell-periodic part of the Bloch wavefunction. Before moving on we introduce the magnetic length corresponding to one flux quantum per unit cell: 
\begin{equation}
    \ell^2_B={|\bm a_1\times\bm a_2|}/({2\pi}).
\end{equation}

 \subsection{Further details}

The properties of the model are encoded in the $C=1$ flat Chern band wavefunction~\cite{wang2021}:
\begin{equation}\label{app:C_1_idealband}
    \psi_{\bk}(\br)=\mathcal N_{\bk} \mathcal B(\br) \Phi_{\bk}(\br),
\end{equation}
where we have introduced the Lowest Landau level wavefunction with one flux quantum in the unit cell $(\a_1,\a_2)$:
\begin{equation}\label{jacobi_LLL}
    \Phi_{\bk}(\br) = 
    \Theta_1(\tilde z -\tilde k,\omega) e^{\frac{\pi}{2\Im\omega}(\tilde z^2-|\tilde z|^2)}e^{\frac{\pi}{2\Im\omega}(\tilde k^2-|\tilde k|^2)} e^{i\bk\cdot\bm t_1\,\tilde z}, 
\end{equation}
with $\omega=e^{2i\pi/3}$, $\mathcal N_{\bk}$ the normalization and $\Theta_1$ the Jacobi Theta function,
\begin{equation}
    \Theta_1(z,\omega)=\sum_{n\in\mathbb Z} e^{i \pi \omega (n+1/2)^2} e^{2 i \pi (z-1/2)(n+1/2)}.
\end{equation}
In Eq.~\eqref{jacobi_LLL}, we have introduced $\tilde z=z/a_1$ ($a_1$ primitive lattice vector in complex notation) and $\tilde k=k/G_2$ ($G_2$ reciprocal lattice vector in complex notation), where $a_2/a_1=\omega$, $G_2=it_1/\ell^2_B$ and $G_1=-\omega G_2$.
The ideal wavefunction~\eqref{app:C_1_idealband} corresponds to the zero mode of Dirac electrons in an effective periodic magnetic field containing one flux quantum per unit cell $B(\br) = 2\pi\hbar / (e|\bm a_1 \times \bm a_2|) + \delta B(\br)$ with $\delta B(\br)$ the periodic part with zero average over the unit cell.
Given $u_{\bk}(\br)=e^{-i\bk\cdot\br}\psi_{\bk}(\br)$, we have: 
\begin{equation}\begin{split}\label{app:form_factors}
\Lambda_{\bk,\bp+\G}=\mathcal N_{\bk}\mathcal N_{\bp}\sum_{\G'}w_{\G'} f^{\bk,\bp}_{\G'+\G},
\end{split}\end{equation}
where we employed the Fourier decomposition: 
\begin{equation}
    |\mathcal B(\br)|^2=\sum_{\G} e^{-i\G\cdot r} w_{\G}, \quad w_{-\G}=w_{\G}^*.
\end{equation}
In Eq.~\eqref{app:form_factors}, we have introduced the overlap between the lowest Landau levels, $f^{\k,\p}_{\G}$.
Exploiting magnetic translation algebra one finds: 
\begin{equation}
    f^{\bk,\bk'}_{\G} = \int_{\rm UC} \frac{d^2r}{\Omega}e^{i(\k-\k'-\G)\cdot\bm r} \Phi^*_{\bk}(\br)\Phi_{\bk'}(\br)=\eta_{\bm G} e^{\frac{i}{2}(\bk+\bk')\times\bm G\ell^2_B}e^{i(\bk\times\bk')\ell^2_B/2} e^{-|\bk-\bk'-\G|^2\ell^2_B/{4}},
\end{equation}
where $\eta_{\G}=(-1)^{n+m+nm}$ is the parity of $\G$ with $\G=n\bm b_1+m\b_2$ and $n,m\in\mathbb Z$. 
Finally, we have the normalization given by: 
\begin{equation}
    \mathcal N^{-2}_{\bk}=\sum_{\G}  \eta_{\G} w_{\G} e^{i\bk\times\G\ell^2_B}e^{-|\G|^2\ell^2_B/{4}}.
\end{equation}
Notice that $|\mathcal B(\br)|=e^{-K(\br)}$~\cite{dong2022diracelectronperiodicmagnetic}, where $K(\br)$ is determined by intra-unit-cell variations of the effective magnetic field: 
\begin{equation}\label{app:B-K_relation}
 B(\br)=\frac{2\pi \hbar}{e|\bm a_1\times\bm a_2|}+\delta B(\br)\implies    \nabla^2K(\br) = \frac{e}{\hbar}\delta B(\br).
\end{equation}
In our work, we consider an ideal Chern band derived from the LLL of a Dirac particle in a $D_6$ symmetric periodic magnetic field that is a sum of the lowest harmonics, \begin{equation}
\label{eq:idealbandmodel}
    \delta B(\br)=\delta B \sum_{i = 1,2,3}\cos(\b_i \cdot \r)
\end{equation} with $\b_1=4\pi/(\sqrt{3}a)(1/2,\sqrt{3}/2)$, $\b_2=-4\pi/(\sqrt{3}a)(1,0)$ and $\b_3=-\b_1-\b_2$. 
Introducing the parametrization: 
\begin{equation}
    K(\r)= -\frac{\sqrt{3}\mathcal K}{4\pi}\sum_{i=1,2,3}\frac{e^{i\b_i\cdot\r}+e^{-i\b_i\cdot\r}}{2}\implies \mathcal K = \frac{|\a_1\times \a_2|\delta B}{\Phi_0},
\end{equation}
where we employed Eq.~\eqref{app:B-K_relation} and $\Phi_0=h/e$ is the flux quantum. Finite values of $\delta B$ introduces nonuniform quantum geometry while keeping the Chern band perfectly flat.

\section{Anomalous Hall metal hole-doping $\nu=1$ and its pairing instability}\label{app_sec:ahm}

An crucial feature of the Hamiltonian \eqref{app:hamiltonian} is its lack of particle-hole symmetry when the  quantum geometry of the ideal band is nonuniform. 
While electron doping near the vacuum at $\nu = 0$ leads to a $c_{\bk}$ quasiparticles with vanishing dispersion~\eqref{app:hamiltonian}, hole doping the $\nu = 1$ state results in a finite bandwidth, directly controlled by inhomogeneities of the quantum geometry. 
The model features an emergent scale in addition to the interaction energy associated with the fluctuations of the quantum geometry. 
For large enough fluctuations, the dispersion becomes dominant and gives rise to an anomalous Hall Fermi liquid of the hole quasiparticles. 
\begin{figure}
    \centering
    \includegraphics[width=0.4\linewidth]{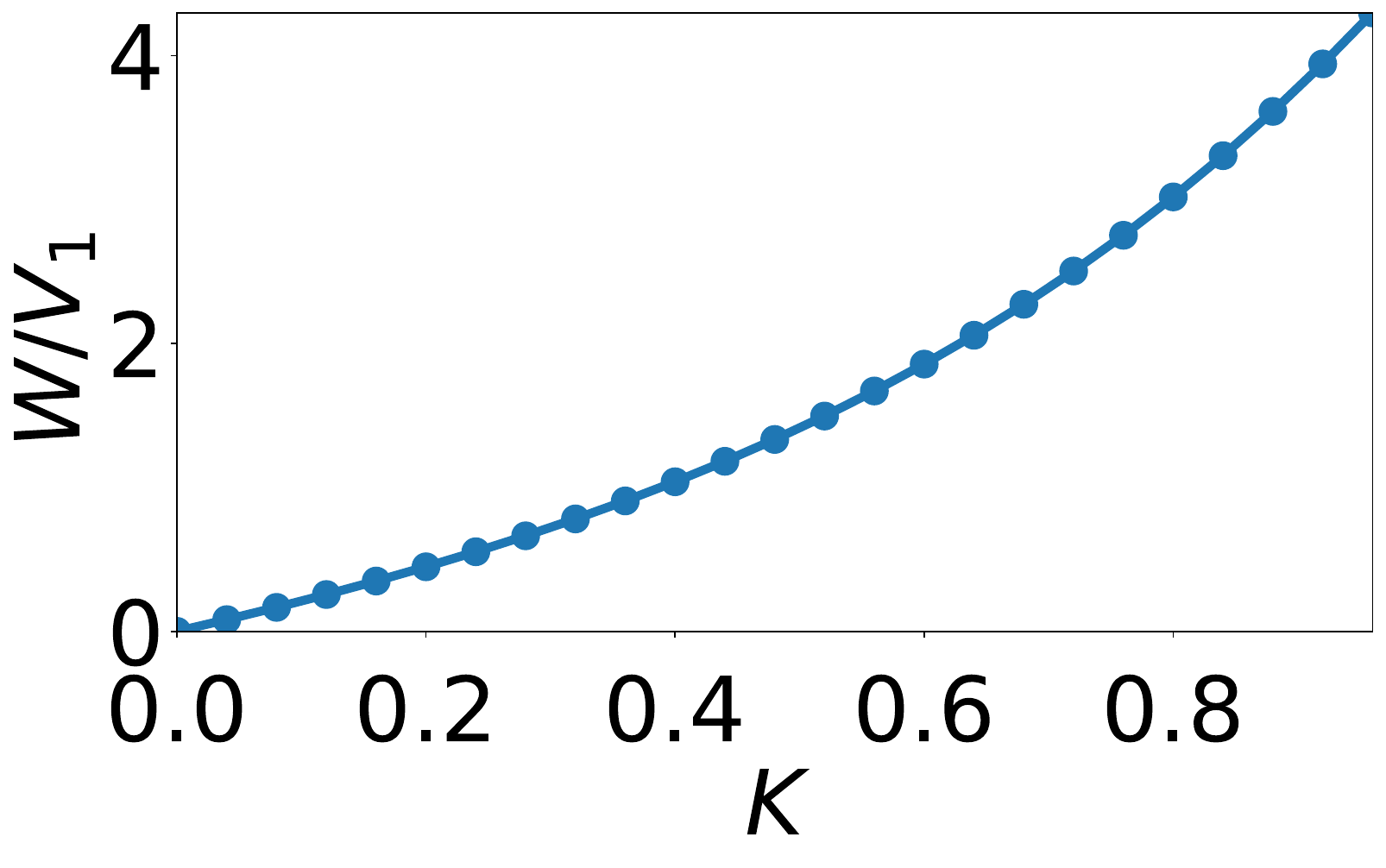}
    \caption{Hole quasiparticles bandwidth $W/V_1$ as a function of $\mathcal K$. For $\mathcal K=0$, in the lowest Landau level limit, the bandwidth vanishes.}
    \label{fig:bandwidth}
\end{figure}

To understand this, we perform a particle-hole (PH) transformation  $c_{\k} \rightarrow d^{\dagger}_{-\k}$ on the Hamiltonian \eqref{hamiltonian} which maps fillings $\nu \rightarrow 1-\nu $ and accordingly, the fully filled ferromagnetic at $\nu=1$,
$ \ket{\Phi_0} = \prod_{\bk\in \rm BZ} c^\dagger_{\bk}\ket{0}$,  becomes vacuum of $d_{\k}$ quasiparticles.
 \begin{equation}\label{ph_transformed_hamiltonian}
    H_d= \mathcal P H \mathcal P^\dagger=\sum_{\bk}\epsilon(\bk) d^\dagger_{\bk}d_{\bk}+ \frac{1}{2A}\sum_{\bk_1\cdots \bk_4}\tilde H_{\bk_1,\bk_2;\bk_3,\bk_4}\, d^\dagger_{\bk_1 }d^\dagger_{\bk_2}d_{\bk_3}d_{\bk_4},
\end{equation}
where $\tilde H_{\bk_1,\bk_2;\bk_3,\bk_4}=H_{-\bk_4,-\bk_3;-\bk_2,-\bk_1}$ up to a constant energy shift. 
The particle-hole transformation applied to the projected density operator gives: 
\begin{equation}\begin{split}
    \mathcal P \bar\rho(\q)\mathcal P^\dagger&=\sum_{\k\in\rm BZ}\braket{u_\k}{u_{\k+\q}}d_{-\k}d_{-\k-\q}^\dagger =N-\sum_{\p\in\rm BZ}\braket{u_{-\p-\q}}{u_{-\p}}d^\dagger_{\p} d_{\p+\q}=N-\sum_{\p\in\rm BZ}\Lambda^*_{-\p,-\p-\q}d^\dagger_{\p} d_{\p+\q}\\
    & = N- \bar n(\q),
\end{split}\end{equation}
with $N$ number of unit cells. Note that under the particle-hole transformation, $\Lambda_{\k,\k+\q} \to \Lambda^*_{-\k,-\k-\q}$, the chirality of the Chern band is inverted—reflecting the fact that electron ($c$) and hole ($d$) quasiparticles experience opposite effective magnetic fields.
Upon performing the particle-hole transformation the quasiparticles $d_{\bk}$ acquire a dispersion relation: 
\begin{equation}\label{quasiparticle_dispersion}
    \epsilon({\bk})=-\Sigma^F_{-\bk}-\Sigma^H_{-\bk},
\end{equation}
with Hartree and Fock self-energies
\begin{equation}\label{self_energies}\begin{split}
    &\Sigma^H_{\bk}=\frac{1}{A}\sum_{\G}V({\G})\rho_{\G}(\bk)\sum_{\k'}\rho_{-\G}(\k'),\\
    &\Sigma^F_{\bk}=-\frac{1}{A}\sum_{\k'}\sum_{\G}V({\k-\k'-\G})|\Lambda_{\bk,\k'+\G}|^2,
\end{split}\end{equation}
$\G=n_1\bm b_1+n_2\bm b_2$ reciprocal lattice vectors and $ \rho_{\G}(\bk)=\Lambda_{\bk,\bk+\G}$.

The bandwidth of the dispersion relation $\epsilon(\bk)$~\eqref{quasiparticle_dispersion} vanishes in the limit of flat quantum geometry, where the PH symmetry is restored. 
On the other hand, for finite $\k$-dependent quantum metric the quasiparticles $d_{\bk}$ acquire a finite dispersion relation. 
The evolution of the bandwidth $W$ as a function of $\mathcal K$ is shown in Fig.~\ref{fig:bandwidth}. 
% Notice that $W$ vanishes for $\mathcal K=0$ and grows monotonously with $\mathcal K$, reflecting the increasing momentum-space inhomogeneity of the quantum geometry.

\begin{figure}
    \centering
    \includegraphics[width=.65\linewidth]{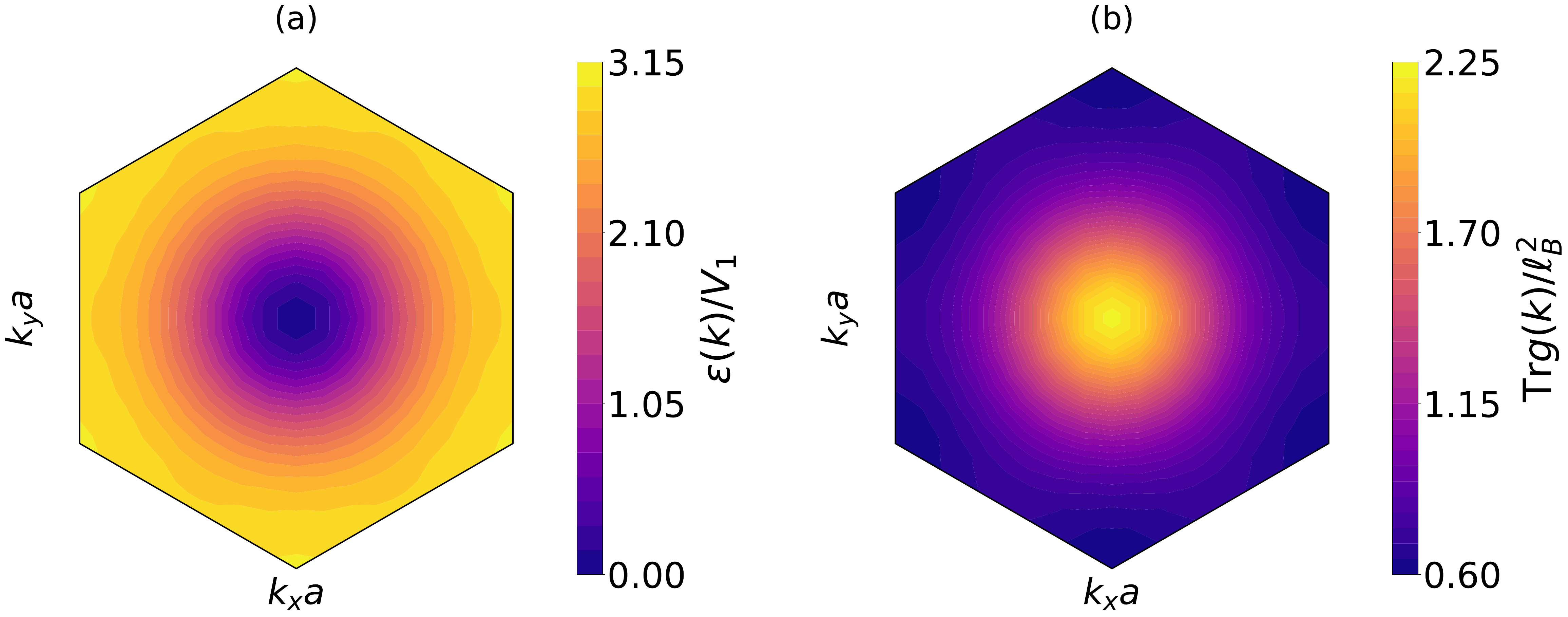}
    \caption{Panels a) and b) show $\epsilon(\bk)/V_1$ and $\Tr g(\bk)/\ell_B^2$ for $\mathcal K=0.8$ in the BZ. The calculations are performed employing $V_1(\br)=(3 V_1 a^4/ 4 \pi)\nabla^2\delta(\br)$. The bandwidth is $W/V_1=3.02$.}
    \label{fig:metric_induced_dispersion}
\end{figure}

In Fig.~\ref{fig:metric_induced_dispersion}(a), we plot the interaction-induced dispersion of the ideal band model \eqref{eq:idealbandmodel}. The dispersion is highly nonuniform and exhibits a minimum at $\gamma$ and maxima at $\kappa$ and $\kappa'$ of the Brillouin zone.
There is a close connection between $\epsilon(\bk)$ with $\Tr g(\bk)$~\cite{Abouelkomsan_2023}, shown in Fig.~\ref{fig:metric_induced_dispersion}(b), which can be understood expanding the absolute value of the form factors $|\Lambda_{\bk,\bk+\bq}|^2\approx 1-q_aq_b g_{ab}(\bk)$, where repeated indeces are summed, in the small momentum limit. 
As a result, expanding for $\bp\approx\bk$ and $\G=0$ the Hartree and Fock dispersion becomes: 
\begin{equation}\label{dispersion}
\epsilon(\bk) \approx \frac{1}{A}\sum_{\bq\in \rm BZ} [V(\q)-V(0)]-\Tr g(\bk)\frac{1}{A}\sum_{\bq\in \rm BZ} \frac{|\bq|^2}{2} V(\q).
    % \Sigma^H_{\bk}\approx \frac{V_{0}}{\Omega},\quad  \Sigma^F_{\bk}\approx-\frac{1}{A}\sum_{\bq} V_{\bq}+\Tr g(\bk)\frac{1}{A}\sum_{\bq} \frac{|\bq|^2}{2} V_{\bq}.
\end{equation}
Thus, the induced dispersion relation is characterized by a $\bk$-dependence closely related to the behavior of $-\Tr g(\bk)$~\cite{Abouelkomsan_2023}. Maxima of $\Tr g(\bk)$ corresponds to minima of $\epsilon(\bk)$ and viceversa. 
Further expanding Eq.~\eqref{dispersion} around $\gamma$, we find: 
\begin{equation}\label{kp_expansion}
    \epsilon(\bk)\approx\frac{k^2}{2m}, \quad  \frac{1}{m}=-\left[\frac{1}{2}\sum_a\partial^2_{k_a}\Tr g(\bk)\right]_{\gamma}\frac{\sum_{\bq\in\rm BZ} |\q|^2 V(\q)}{2A}.
\end{equation}
For large enough Berry curvature fluctuations, owing to the small effective mass $m$, the doped quasiparticles form an anomalous Hall metal with Fermi energy $\epsilon_F = \frac{k^2_F}{2 m }$ $(\pi k_F^2=\delta|\b_1\times\b_2|)$ and residual interaction $\tilde H_{\bk_1,\bk_2;\bk_3,\bk_4}$. 
This Fermi liquid exhibits anomalous Hall conductivity, wherein local charge fluctuations induce variations in orbital magnetization, and conversely, magnetic field perturbations give rise to charge responses~\cite{Haldane2004}. 
This interplay between charge and magnetization reflects the underlying chirality encoded in the form factors $\Lambda_{\bk,\bp}$, with implications for interaction-induced phenomena such as superconducting instabilities in the anomalous Hall metal.

\section{Long-wavelength theory of the anomalous Hall metal}\label{app_sec:long_wavelength}

Within the anomalous Hall metal phase, at low-doping, we perform a $\bk\cdot\bp$ expansion retaining the residual interaction between the quasiparticles. 
First, we notice that for small $\delta$, $ak_F\ll1$, we can neglect umklapp processes involving finite reciprocal lattice vectors $\G$, as well as expand the dispersion relation around $\gamma$. 
The resulting low-energy theory reads: 
\begin{equation}\label{app:low_energy_model}
    \mathcal H =     \mathcal H=\sum_{\bk}\frac{\bk^2}{2m_*} d^\dagger_{\bk} d_{\bk}+\frac{1}{2A}\sum_{\bq\bk\bk'} V(\q)\Lambda^*_{\k,\k+\q}\Lambda^*_{\k',\k'-\q}d^\dagger_{\bk}d^\dagger_{\bk'}d_{\bk'-\bq}d_{\bk+\bq},
\end{equation}
where 
\begin{equation}\label{kp_expansion_formfactors}
    \Lambda_{\bk,\bk'}\approx 1+i\ell^2 \bk\times\bk'/2-\ell^2(\bk-\bk')^2/4\approx e^{i\ell^2(\bk\times\bk')/2-\ell^2|\bk-\bk'|^2/4}.
\end{equation} 
In the latter expression, $\ell^2$ is given by the quantum metric of the ideal wavefunction at $\Gamma$: 
\begin{equation}
    \ell^2=\Tr g(0)=\frac{|\a_1\times \a_2|}{2\pi}\left(1-\frac{\mathcal N^2_0 }{2}\sum_{\G} \ell^2_B  \G^2 w_{\G} \eta_{\G}e^{-\G^2\ell^2_B/4}\right).
\end{equation}
Derivatives of the Berry curvature and quantum metric appear as higher corrections to Eq.~\eqref{kp_expansion_formfactors}.

We consider the  the leading pseudopotential term beyond a contact interaction: $V_1=v_1\nabla^2 \delta(\br)$ and $v_1=3 V_1 a^4/ 4 \pi$. 
For small momentum transfer, the projected $V_1$-interaction in the particle-particle channel reads: 
\begin{equation}\begin{split}
    \Gamma_{\k,\k'}&=\frac{v_1}{2} \left(-(\k-\k')^2\Lambda_{\k',\k}\Lambda_{-\k',-\k}  +(\k+\k')^2\Lambda_{-\k',\k}\Lambda_{\k',-\k} \right)\\
    &=v_1 e^{-\ell^2\frac{k^2+k'^2}{2}}\left[2\k\cdot\k'\sum_{j=0}^{\infty}\frac{(\ell^2 k_+ k'_-)^{2j}}{2j!}-(k^2+k'^2)\sum_{j=0}^{\infty}\frac{(\ell^2 k_+ k'_-)^{2j+1}}{(2j+1)!}\right],
\end{split}\end{equation}
where we have expanded the form factors to small momenta and $\ell k<1$ and $k_\pm=k_{x}\pm ik_y$.
Including terms up to $f$-wave angular momentum we recover the expression in the main text. 
The interaction develops a finite and repulsive components in angular momentum channels $L^z=\cdots,-(2n+1),\cdots,-1,+1$, while channels $L^z>+1$ remain unaffected. We define the interaction projected in the angular components $L^z=m$ as: 
\begin{equation}\label{projected_interaction}
    \Gamma^{m}_{k,k'}=\int^{2\pi}_0\frac{d\phi}{2\pi} e^{im\phi} \Gamma_{\k,\k'},
\end{equation}
where $\phi=\theta_{\k}-\theta_{\k'}$.  
Finally, we observe that the chirality is inverted upon performing the particle-hole transformation, going back to the original formulation, $d_\k\to c^\dagger_{-\k}$ the particle-particle interaction transforms as $\Gamma_{\k,\bk'}\to \Gamma^*_{\k,\k'}$. This implies that the $\nu = 1/3$ Laughlin state for holes exhibits opposite chirality compared to its electron counterpart.

The lowest angular momentum channel with vanishing bare repulsion is $L^z = +3$.
A weak attraction can therefore stabilize pairing in this chiral channel. In what follows, we examine how screening induces such pairing.

\subsection{Pairing Induced by Overscreened Interactions}

\begin{figure}
    \centering
    \includegraphics[width=0.5\linewidth]{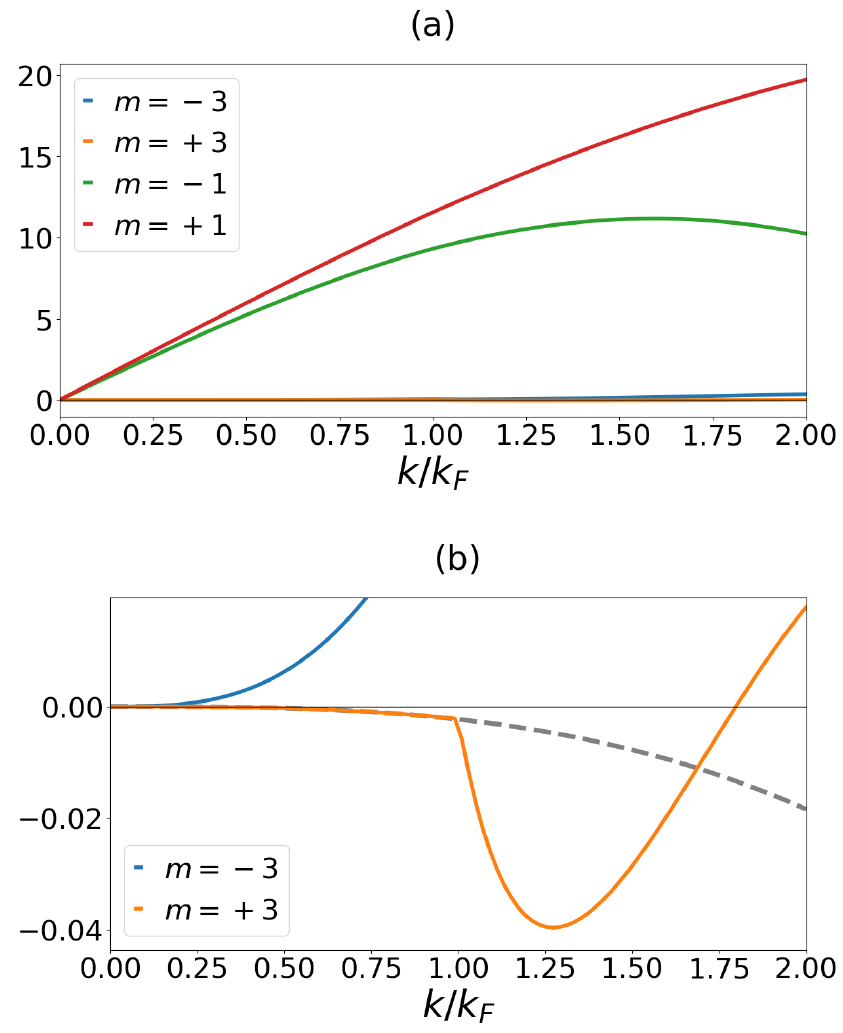}
    \caption{Panel (a): Particle-particle interaction $\tilde \Gamma^m_{k,k'=k_F}/v_1$ for different angular momentum as a function of $k/k_F$. The dominant components, $L^z = \pm1$, are strongly repulsive and split due to time-reversal symmetry breaking. Panel (b) shows a zoom in the higher angular momentum components $m=\pm3$. The dashed line displays the small momentum result in Eq.~\eqref{small_momentum_apprx}. 
    Notice that the attraction on the Fermi surface, $\tilde \Gamma^{+3}_{k_F,k_F}<0$, has a purely geometric origin and vanishes for $\ell=0$.
    Parameters used in the calculations are $\ell k_F=0.32$, corresponding to $\delta=.05$}
    \label{fig:weak_coupling}
\end{figure}

In the following, we consider the screening of the interaction as a possible mechanism for pairing between $d_{\k}$ quasiparticles. 
Within the random phase approximation, we sum an infinite series of particle-hole bubble diagrams, effectively replacing the bare interaction $V(\q)=-v_1\q^2$ is replaced by: 
\begin{equation}\label{dressed_interaction}
    \tilde V(\q)=\frac{V(\q)}{1+V(\q)\chi(\q,0)},
\end{equation}
where we have introduced the static density-density response function: 
\begin{equation}
\chi(\bq,i\Omega=0)=-\frac{T}{A}\sum_{\bp}\sum_{i\omega_n}\frac{|\Lambda_{\p,\p+\q}|^2}{(i\omega_n-\xi_{\p+\q})(i\omega_n-\xi_{\p})}=4m\int^{k_F}\frac{d^2\p}{(2\pi)^2}\frac{|\Lambda_{\bp,\bp+\bq}|^2}{q^2+2\bp\cdot\bq},    
\end{equation}
$\xi_{\k}=\k^2/2m-\mu$ with $\mu$ chemical potential and Matsubara frequency $\omega_n=(2n+1)\pi k_BT$. The last expression is obtained in the zero-temperature limit. 

For $\ell k_F\ll 1$, when the characteristic wavefunction variation scale $1/\ell$ is much larger than the Fermi momentum, i.e., $1/\ell \gg k_F$, we have $|\Lambda_{\mathbf{p}, \mathbf{k}}| \approx 1$, and the resulting response function reduces to the density-density response of the 2DEG~\cite{Giuliani_Vignale_2005}: 
\begin{equation}
    \chi(\q,0) \approx D(0)\left[1-\theta\left(\frac{q}{k_F}-2\right)\frac{\sqrt{(q/k_F)^2-4}}{q/k_F}\right],
\end{equation}
where $D(0)=m/(2\pi)$ is the density of states.

Quantum geometric correction gives rise to a $\q$-dependency of the response function of $\chi(\q)$ for $q\le 2k_F$. 
As a result, $\chi(\q)$  develops finite angular momentum components for momentum transfer below $2k_F$. 
Introducing the screened interaction in the pair scattering amplitude, we find: 
\begin{equation}
    \tilde \Gamma_{\k,\k'}=\frac{1}{2}\left(\tilde V(\k-\k')\Lambda_{\k',\k}\Lambda_{-\k',-\k} -\tilde V(\k+\k')\Lambda_{-\k',\k}\Lambda_{\k',-\k} \right),
\end{equation}
where $\tilde V(\q)$ is given in Eq.~\eqref{dressed_interaction}. Eq.~\eqref{projected_interaction} is employed to decompose $\tilde \Gamma_{\k,\k'}$ in angular momentum sectors.

Figure~\ref{fig:weak_coupling}(a) shows $\tilde \Gamma^{m}_{k, k' = k_F}$ as a function of $k$, while panel (b) focuses on the low-energy sector, highlighting angular momenta $m = \pm 3$. Unlike the $L^z = -3, -1, 1$ channels, which are dominated by bare repulsion, $L^z=3$ develops a net attraction. 
This can be readily understood expanding the expression for small momentum transfer $\q\le 2k_F$ and for small $V_1$: $\tilde \Gamma_{\k,\k'}=\Gamma_{\k,\k'}+\delta\Gamma_{\k,\k'}$ where to second order in $V_1$ we have: 
\begin{equation}
    \delta\Gamma_{\k,\k'}=-\frac{V(\k-\k')\chi(\k-\k')V(\k-\k')\Lambda_{\k',\k}\Lambda_{-\k',-\k}-V(\k+\k')\chi(\k+\k')V(\k+\k')\Lambda_{-\k',\k}\Lambda_{\k',-\k}}{2}.
\end{equation}
Expanding the expression for small $\ell k_F$ and projecting it in the $L^z=+3$ angular momentum channel, we find the leading contribution in $\ell k_F$: 
\begin{equation}\label{small_momentum_apprx}
    \tilde \Gamma_{\k,\k'}\approx -v_1^2 D(0)\ell^2(k_-k_+')^3,
\end{equation}
resulting in the low-energy attractive interaction: 
\begin{equation}
    \mathcal H^{\rm int}_{\k,\k'} = - v_1^2D(0)\ell^2 (\mathcal F^+_\k)^\dagger \mathcal F^+_{\k'}.
\end{equation}
Here, $\mathcal F^+_{\k}=k^3_+d_{-\bk}d_{\k}$ denotes the chiral $f$-wave pair operator with relative momentum $\k$ and vanihsing center of mass momentum, and the coupling constant $v_1^2D(0)\ell^2$. 

The resulting Bogoliubov-de Gennes Hamiltonian reads: 
\begin{equation}
    \mathcal H_{\rm BdG}=
    % \frac{1}{2}\sum_{\k}\left[\xi_{\k}d^\dagger_{\k}d_{\k}-\xi_{\k} d_{-\k}d^\dagger_{-\k}\right]+\frac{\Delta}{2}\sum_{\k}\left[k_-^3 d^\dagger_{\k}d^\dagger_{-\k}+h.c.\right],
    \frac{1}{2}\sum_{\k}\Psi^\dagger_{\k}\begin{pmatrix}\xi_{\k} & \Delta k^3_-\\ 
    \Delta k^3_+ & -\xi_{\k}\end{pmatrix}\Psi_{\k},
\end{equation}
where we have introduced the constant:
\begin{equation}
    \Delta = - v_1^2D(0)\ell^2 \sum_{\k} k^3_+ \langle d_{-\bk}d_{\k}\rangle,
\end{equation}
and the Nambu spinor $\Psi_{\k}=(d_{\k},\,d^\dagger_{-\k})^T$. The Chern number of the BdG Hamiltonian can be directly obtained by noting that the in-plane component of the vector exhibits a winding of 3, while the $z$-component increases with increasing $k$, resulting in a Chern number $C = +3$.

Next, we introduce the quasiparticle operators $\alpha_{\k}=U_{\k} d_{\k}-V_{\k} d^\dagger_{-\k}$, with $|U_\k|^2+|V_\k|^2=1$, such that $\{\alpha_{\k},\alpha^\dagger_{\k'}\}=\delta_{\k,\k'}$. 
The BCS ground state, defined as the vacuum of the quasiparticles $\alpha_{\k}\ket{\Psi}=0$, reads: 
\begin{equation}\label{bcs_wavefunction_1}
    \ket{\Psi} = \prod'_{\k} \alpha_{\k}\alpha_{-\k}\ket{0}= \prod'_{\k}\left[U_{\k} + V_{\k} d^\dagger_{\k} d^\dagger_{-\k}\right]\ket{0}.
\end{equation}
Here, the prime on the product indicates that each distinct pair $(\k, -\k)$ is counted only once, and the last equality holds up to a normalization constant.

The coefficients $U_{\k}$ and $V_{\k}$ are obtained solving the quasiparticle equation $[\alpha_{\k},\mathcal H_{\rm BdG}]=E(\k)\alpha_{\k}$, which yields: 
\begin{equation}
    \begin{split}
        & E(\k)U_{\k}=\xi_{\k}U_{\k}-\Delta(\k)^*V_{\k},\\
        & E(\k)V_{\k}=-\xi_{\k}V_{\k}-\Delta(\k)U_{\k},
    \end{split}
\end{equation}
where $\Delta(\k)=\Delta k^3_-$.
By solving these equations one readily finds: 
\begin{equation}
    |U_{\k}|^2=\frac{1}{2}+\frac{\xi_{\k}}{2E(\k)},\quad  |V_{\k}|^2=\frac{1}{2}-\frac{\xi_{\k}}{2E(\k)},\quad V_{\k}=-\frac{E(\k)-\xi_{\k}}{\Delta(\k)^*}U_{\k}, \quad E(\k)=\sqrt{\xi^2_{\k}+|\Delta(\k)|^2}.
\end{equation}
Due to the Pauli exclusion principle, $\left(d^\dagger_{\k}\right)^2=0$, the BCS ground state can be expressed as: 
\begin{equation}\label{bcs_wavefunction_2}
    \ket{\Psi}=\prod_\k |U_\k|^{1/2}\exp\left(\sum_{\k}G_{\k} d^\dagger_{\k}d^\dagger_{-\k}/2\right)\ket{0},
\end{equation}
where the product is now extended to all $\k$-points, which also accounts for the square root in $|U_{\k}|$ and the prefactor $1/2$ in the exponent. We have introduced $G_{\k} = \frac{V_{\k}}{U_{\k}} = -\frac{E(\k) - \xi_{\k}}{\Delta(\k)^*}$.
The operator at exponent of Eq.~\eqref{bcs_wavefunction_2}: 
\begin{equation}
    b^\dagger_{\q=0} = \frac{1}{2}\sum_{\k}G_{\k} d^\dagger_{\k}d^\dagger_{-\k},
\end{equation}
defines the two-body Cooper pair at zero momentum, which is macroscopically occupied in the ground state.
The real-space Cooper pair wavefunction can be obtained as: 
\begin{equation}
    G(\r)=\mel{0}{\psi(\r)\psi(\r'=0)b^\dagger_{\q=0}}{0}=\frac{1}{A}\sum_{\p}e^{i\p\cdot\r} G_{\p}=\frac{1}{\Delta }\int\frac{d^2\p}{(2\pi)^2}e^{i\p\cdot\r} \frac{\xi_{\p}-E(\p)}{(p_x+ip_y)^3},
\end{equation}
where $\psi(\r)$ is a fermionic field annihilation operator at $\r$ and in the last step we took the continuum limit.
From the latter expression we readily realize that: 
\begin{equation}
    G(R_\theta \r) = e^{-3i\theta}G(\br),
\end{equation}
consistent with the angular momentum behavior with $L^z=+3$, in our notation~\eqref{projected_interaction}. Moreover, performing the angular integral we find; 
\begin{equation}
    G(\r) = i\Delta\frac{e^{-3i\phi}}{2\pi}\int^{\infty}_0 dp\frac{p^4}{(p^2/2m-\mu)+\sqrt{\Delta^2p^6+(p^2/2m-\mu)^2}}J_3(pr),
\end{equation}
where $\phi$ is the polar angle of $\r$ measured from the $x$-axis and $J_3(pr)$ is the Bessel function of the first kind. 
The latter integral is dominated by the Fermi surface contribution $p_F=\sqrt{2m\mu}$ and, for large distance $k_Fr\gg 1$, we find $G(\r)\propto1/(x+iy)^3$.

\subsubsection{Linearised gap equation}

The linearized gap equation reads: 
\begin{equation}
    \Delta(\bp)=-\frac T A\sum_{i\omega_n}\sum_{\bk}\tilde \Gamma_{\bp,\bk}G(\bk,i\omega_n) \Delta(\bk)G(-\bk,-i\omega_n),
\end{equation}
where the pairing gap reads
\begin{equation}
\Delta(\p)=-\sum_{\k}\tilde \Gamma_{\p,\k}\langle d_{-\k}d_{\k} \rangle, 
\end{equation} 
$G(p)=(i\omega_n-\xi_\p)^{-1}$ the quasiparticle Green's function, $\xi_\p=\p^2/2m-\mu$ and $\omega_n=(2n+1)\pi k_BT$.
Performing the integral over Matsubara frequencies we find: 
\begin{equation}
    \Delta(\p)=-\frac{1}{A}\sum_{\bk}\tilde\Gamma_{\p,\k}\frac{\tanh\xi_{\k}/(2k_B T)}{2\xi_{\k}}\Delta(\k).
\end{equation}
The latter expression can be expressed as an eigenvalue equation for the pairing kernel $\mathcal M_{\k,\k'}=-I_{\k}\tilde\Gamma_{\k,\k'}I_{\k'}/A$ with $I_{\k}=\sqrt{\frac{\tanh(\xi_{\k}/(2k_B T))}{2\xi_{\k}}}$ and eigenvectors $\phi(\k)=\Delta(\k)I_{\k}$. 
\begin{figure}
    \centering
    \includegraphics[width=0.6\linewidth]{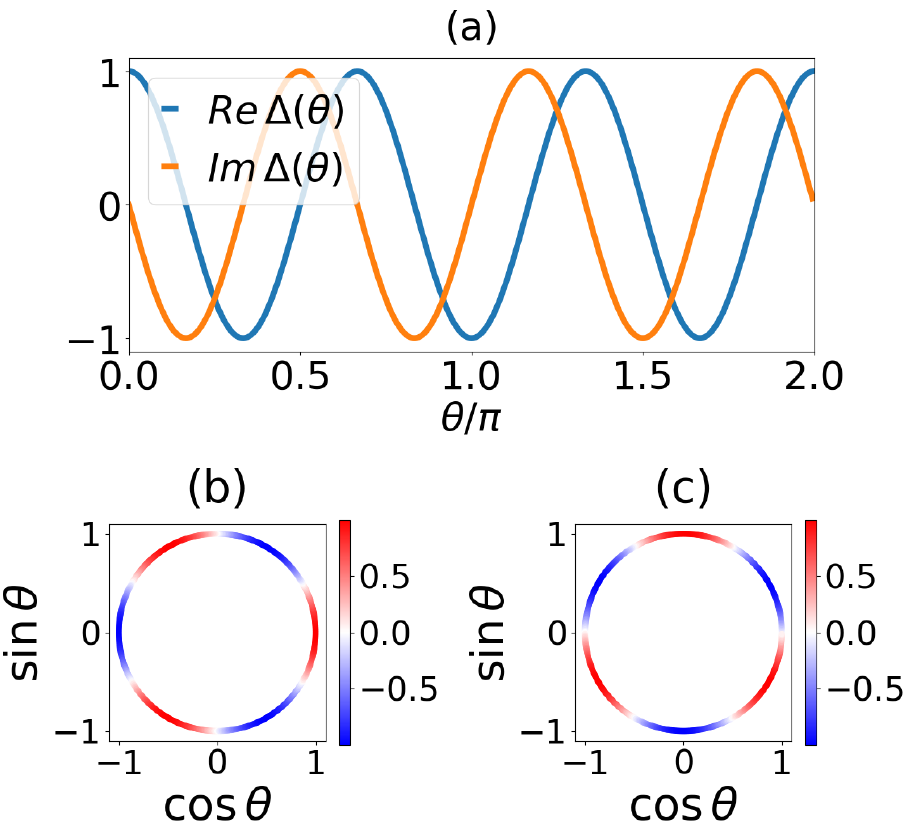}
    \caption{Panel (a) shows the evolution of the pairing gap $\Delta(\theta)$ along the Fermi surface for the leading superconducting instability. Panels (b) and (c) provide two-dimensional visualizations on the Fermi surface. }
    \label{fig:pairing_gap}
\end{figure}

Focusing on pairing around the Fermi surface, we linearize the dispersion around the Fermi surface $\xi_{\k}\approx v_F k_\perp$ with $k_\perp$ perpendicular to the Fermi surface and $v_F=k_F/m$. 
Introducing $k_\parallel$ along the Fermi surface, we replace $\int d^2\k/(2\pi)^2=\int dk_\perp dk_\parallel/(2\pi)^2$ and the integral over $k_\perp$ gives a factor $\log(W/T)/v_F$ with $W$ energy cutoff given by the interaction-induced bandwidth.  
Superconducting instabilities are obtained by evaluating the matrix~\cite{Ghazaryan_2021} on an equidistant grid of points along the Fermi surface~\cite{Ghazaryan_2021}: 
\begin{equation}
    \tilde{\mathcal M}_{\k,\k'}=-\frac{1}{(2\pi)^2}\sqrt{\frac{\Delta k\Delta k'}{v_F^2}}\tilde \Gamma_{\k,\k'},
\end{equation} 
where $\Delta k$ is the distance between two points on the Fermi surface. The eigenvector of $\tilde{\mathcal M}_{\k,\k'}$, denoted $\phi(\theta)$, with the largest eigenvalue gives the leading superconducting instability. Notice that the two are related up to a multiplicative factor which is $\k$ independent for parabolic bands, $\Delta(\theta)=\phi(\theta)\sqrt{v_F/\Delta k}$.

Fig.~\ref{fig:pairing_gap} shows the angular dependence of the paring gap with largest eigenvalue along the Fermi surface. 
The angular dependence exhibits maxima of $\Re\Delta$ at $\theta=2\pi n/3$ ($n=0,1,2$) and of $\Im\Delta$ at $\pi/2+2\pi n/3$ ($n=0,1,2$) characteristic of the chiral $f$-wave $\Delta(\theta)\propto e^{-3i\theta}$. 
This angular profile aligns with the perturbative analysis presented in the previous section, where we found $\Delta(\k)\propto k^3_-$.

\section{Additional ED results}\label{app_sec:ed}

In this section, we provide additional numerical results in support to our theory. 

\begin{figure}
    \centering
\includegraphics[width= \linewidth]{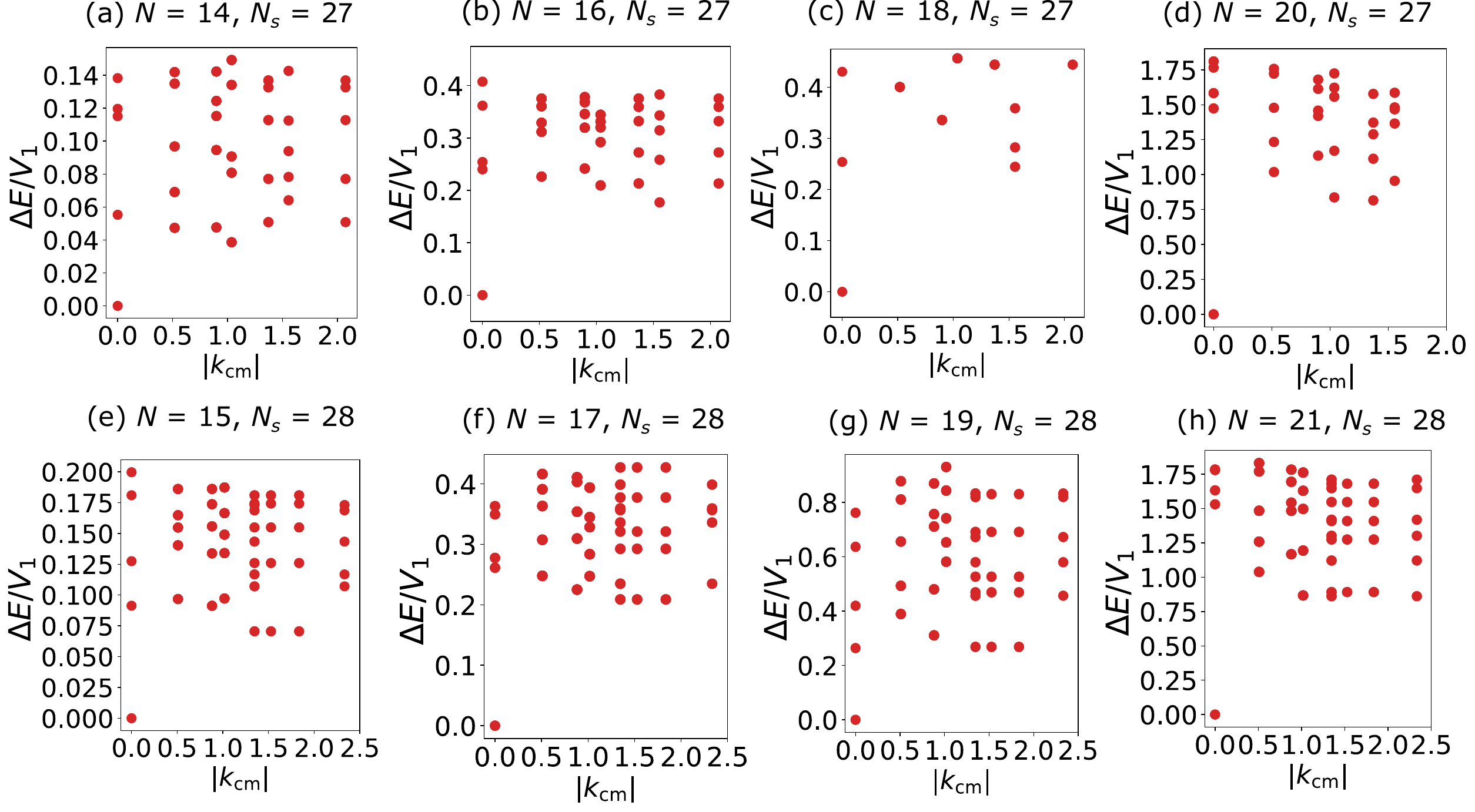}
    \caption{Many-body spectrum at $\mathcal K = 0.8$ evaluated for different values of $N$ for $N_s = 27$ and $N_s = 28$. }
    \label{fig:EDplots}
\end{figure}

\begin{figure}
    \centering
    \includegraphics[width=\linewidth]{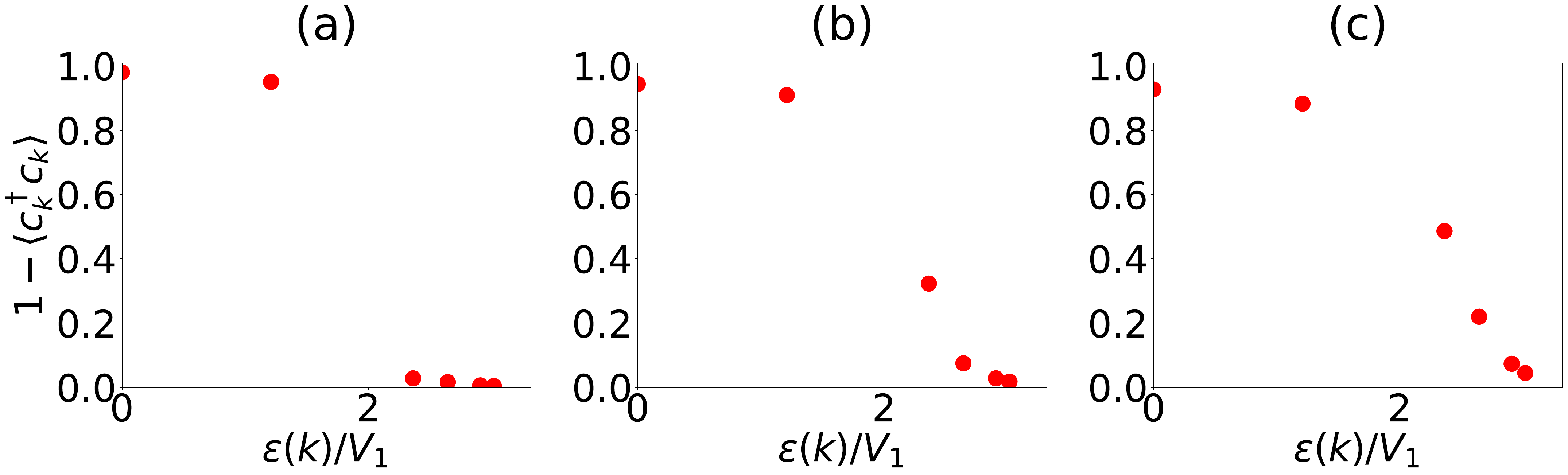}
    \caption{Panels (a), (b) and (c) show the hole occupation number $1-\langle c^\dagger_{\k} c_{\k}\rangle$ as a function of the interaction-induced dispersion $\epsilon(\k)$ for $N=20$, $N=18$ and $N=16$, respectively, $N_s=27$ and $\mathcal K=0.8$. }
    \label{fig:nk_epsk}
\end{figure}

\subsection{Many-body spectrum}

\begin{figure}
    \centering
    \includegraphics[width=0.5\linewidth]{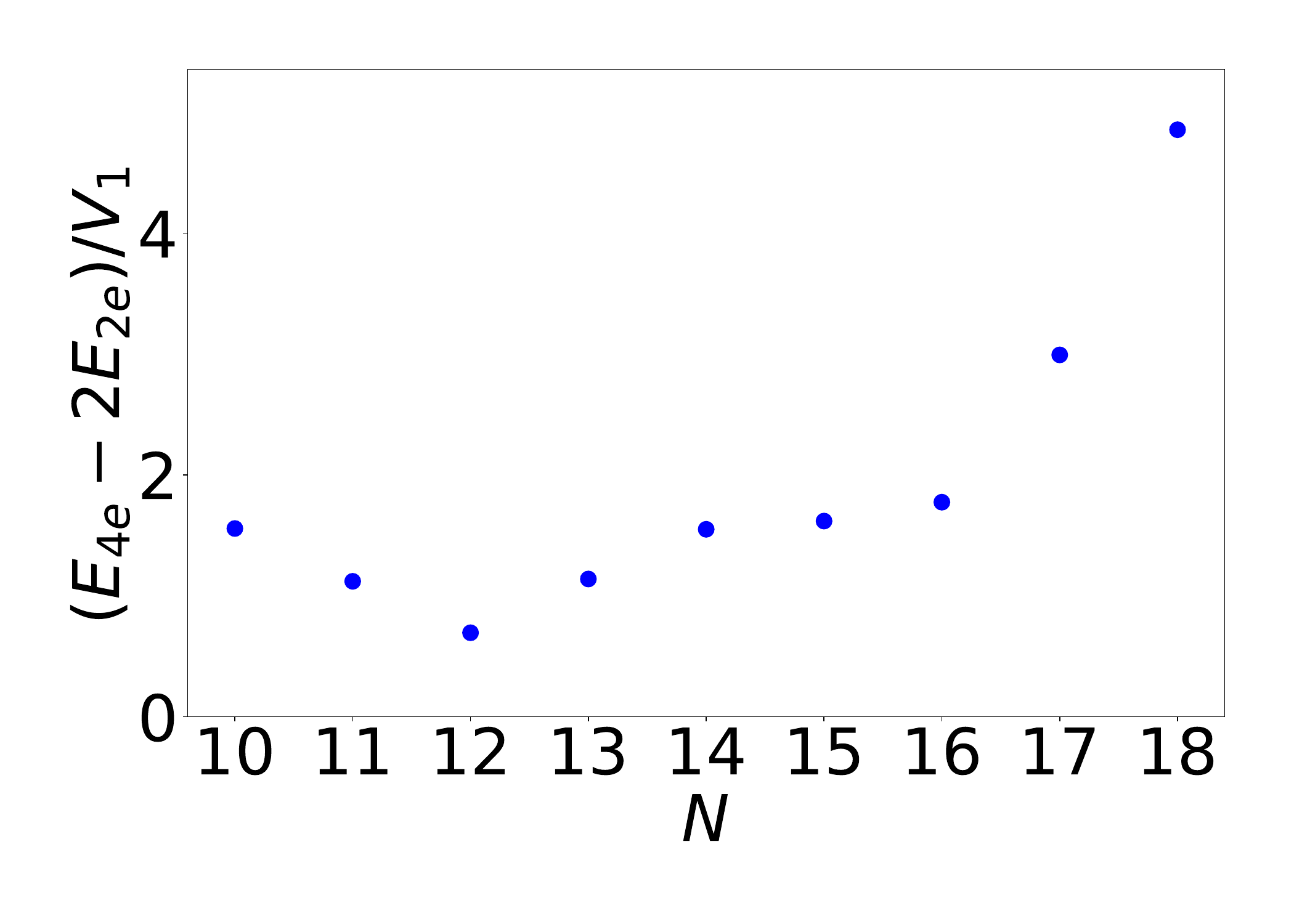}
    \caption{Energy of a charge $4e$ excitation minus twice the energy of two charge $2e$ for $N_s=27$ and $\mathcal K=0.8$. }
    \label{fig:4evs2e_energy}
\end{figure}

Here, we show the many-body spectrum at the $N$ values that give rise to a negative binding energy (Fig. 4 in the main text). As shown in Fig. \ref{fig:EDplots}, the many-body ground state lies at center of mass (COM) momentum $|\k_{\rm cm}| = 0$ consistent with a weak pairing superconducting state as discussed in the main text. 
% \textcolor{red}
{At filling factor $2/3$, the binding energy is $|E_b|/V_1=0.18$ for $N_s=27$ and $|E_b|/V_1=0.13$ for $N_s=21$, with corresponding average kinetic energies per hole $E_K/V_1=1.44$ and $1.32$. Consequently, the binding energy per hole relative to the kinetic energy is nearly constant: $|E_b|/(2E_K)=0.05$ for $N_s=21$ and $0.06$ for $N_s=27$.}
% \textcolor{red}
{ Fig.~\ref{fig:4evs2e_energy} compares the energy of four doped carriers, $E_{4e}=E(N+4)-E(N)$, with twice the energy of two charge-$2e$ excitations, $E_{2e}=E(N+2)-E(N)$. We find $E_{4e}>2E_{2e}$, indicating that once binding develops, the charge-$2e$ excitation is the lowest-energy charged excitation.}

\begin{figure}
    \centering
    \includegraphics[width=0.5\linewidth]{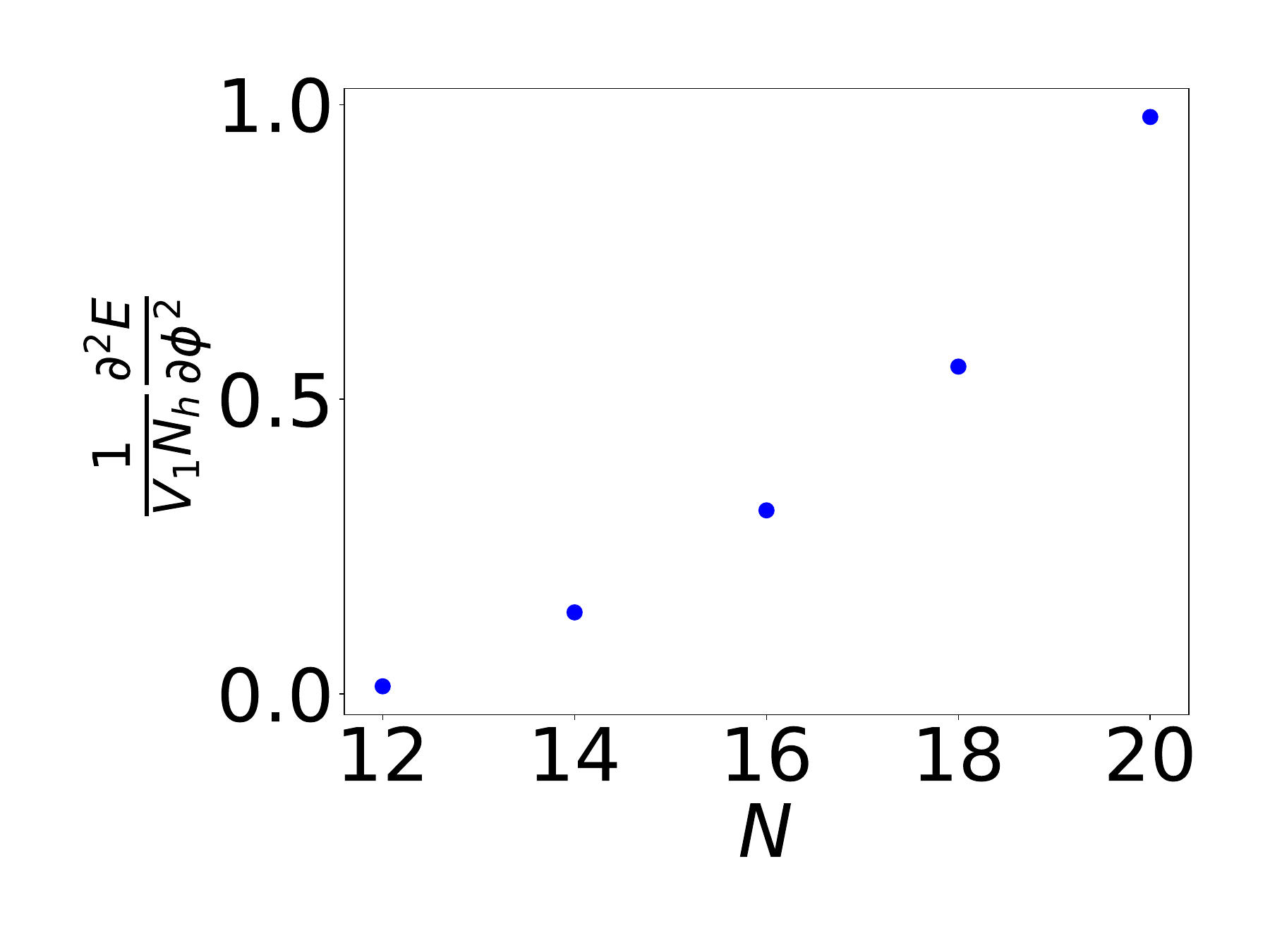}
    \caption{Ground state phase rigidity density versus particle number for $N_s = 27$ and $\mathcal{K} = 0.8$.}
    \label{fig:stiffness}
\end{figure}

% \textcolor{red}
{In the filling-factor regime $N = 14, 16,$ and $18$, where electron pairs exhibit binding, the ground state displays a finite stiffness, shown in Fig.~\ref{fig:stiffness}. This stiffness is computed as the second derivative of the ground-state energy per hole with respect to the flux $\phi$ threading one of the torus handles~\cite{Scalapino1992,Scalapino1993}: $\frac{1}{N_h}\left.\frac{\partial^2 E(\phi)}{\partial \phi^2}\right|_{\phi=0}$.} 

\subsection{Quasiparticle occupation number}

Fig.~\ref{fig:nk_epsk} shows the hole occupation number $1-\langle c^\dagger_{\k} c_{\k}\rangle$ as a function of $\epsilon(\k)$ for different values of $N$ giving rise to a finite binding energy. 
These results further support the Fermi liquid-like nature of the ground state developed at $\mathcal K=0.8$ around $\nu=2/3$.

\end{document}